\documentclass[structabstract]{aa}  
\usepackage{graphicx}
\usepackage{longtable}
\usepackage{txfonts}
\usepackage{natbib}
\bibpunct{(}{)}{;}{a}{}{,}
\begin{document}
   \title{Revisiting the low-luminosity galaxy population of the NGC 5846 group with SDSS}
   \titlerunning {Low-luminosity galaxies in the NGC 5846 group}

   \author{P. Eigenthaler \and W.W. Zeilinger}
   \institute{Institut f\"ur Astronomie, Universit\"at Wien, T\"urkenschanzstra\ss e 17, A-1180 Vienna\\ 
    \email{eigenthaler@astro.univie.ac.at}}

   \date{Received xxx; accepted xxx}

 
  \abstract
   {Low-luminosity   galaxies   are  known   to  outnumber  the   bright   galaxy population    in   poor  groups  and   clusters
   of   galaxies. Yet,   the  investigation    of   low-luminosity galaxy   populations outside     the  Local     Group remains    rare   and     the
   dependence      on different     group environments   is  still     poorly  understood.   Previous  investigations   revealed photometric  scaling
   relations for  early-type dwarfs and a  strong dependence of morphology with environment.}
   {The present   study  aims   to  analyse   the photometric    and   spectroscopic  properties  of    the low-luminosity  galaxy population   in the
   nearby, well-evolved   and    early-type   dominated NGC~5846    group  of    galaxies.  It  is     the     third     most    massive aggregate  of
   early-type   galaxies      after   the          Virgo     and         Fornax     clusters            in        the           local        universe.
   Photometric scaling relations  and   the  distribution    of   morphological  types  as well as  the characteristics of  emission-line galaxies are
   investigated.}
   {Spectroscopically selected low-luminosity  group  members from   the  Sloan  Digital  Sky  Survey with   $cz<3000$ km   s$^{
   -1}$   within  a radius  of  $2\degr=0.91$ Mpc    around  NGC~5846  are   analysed.   Surface    brightness  profiles    of  early-type    galaxies
   are   fit  by    a S\'{e}rsic  model $\propto$  $r^{1/n}$. Star  formation  rates, oxygen abundances  and emission characteristics  are  determined
   for  emission-line galaxies.}
   {Seven new group members    showing     no     entry    in     previous  catalogues are  identified    in     the outer ($>$80     arcmin) parts of
   the  system. Several   photometric scaling  relations for  dEs as  well  as  the morphology-density  relation for  dwarf galaxies  are  reproduced.
   Moreover, the correlation between host  and  satellite morphologies   in poor groups  of  galaxies is   confirmed. Nucleated dwarfs  are  found  to
   be located in  the   vicinity  to the  brightest  ellipticals  in  the group.  Only two faint  galaxies show  fine structure. Emission-line  dwarfs
   show no interaction induced activity.}
   {}


   \maketitle
%

\section{Introduction}
Redshift surveys   have  shown    that  most    galaxies in     the  nearby    universe are       located in   poor   groups      \citep{tullyfisher}.
These   aggregates     are   not only   made  up  of   the  ordinary   bright   Hubble  types  but  consist  also  of  a low-luminosity  dwarf  galaxy
population outnumbering their brighter counterparts by far.  Previous  observations studying  the  distribution and  morphological properties of  such
low-luminosity    galaxy     populations   have  mainly           focused            on        large-scale      structures,  e.g.   the          Virgo
and Fornax       clusters  \citep{sandagebinggeli,    ferguson}  representative of    high-density environments.   Complementary, poor   groups  serve
as ideal  laboratory   to    study   the   environmental   dependence  of    low-luminosity    galaxy   populations   in   a  low-density  environment
with galaxy  interactions,     mergers,  and  the    coalescence    of  individual   galaxies      as   the      driving physical processes.  However,
detailed photometry and spectroscopy of low-luminosity galaxies is challenging. Indeed, only     the    Local      Group and      a     few      other
nearby  galaxy  aggregates   have       been   investigated      in     detail to   analyse        the   connection   between    low-luminosity 
galaxies     to their environment \citep{karachentsev,cote,jerjen,gruetzbauch}. 
Because   $\Lambda$CDM  cosmology    suggests a   hierarchical  growth   of galaxies   in the   course  of   mergers, it is  obvious to  classify poor
groups  based on their                 optical         morphologies \citep{zabludoff}.    Spiral     dominated   aggregates   represent      unevolved
systems  in  this scenario  while     aggregates   with     tidal    signatures   and   the  formation     of   tidal     dwarf  galaxies    mark   an
intermediate    stage  \citep{temporin}.  Groups     dominated   by   early-type    morphologies  are in      an    advanced   phase  of  coalescence.
Dynamically,  these well-evolved aggregates     are  more    likely  to      show   virialised systems compared   to   groups with a   higher   spiral
fraction.   Moreover, the  dwarf-to-giant     ratio   (DGR) is      expected   to   increase      with  time as        bright,  massive      galaxies
are more  affected  by    mergers than    their  faint            counterparts  \citep{zabludoffmulchaey}.     The number   of   faint   galaxies   in
groups   and clusters    is   puzzling,  however. $\Lambda$CDM  cosmology   suggests  more dwarfs   than observed, an  issue widely   referred to   as
the   missing satellite  problem  \citep{klypin}. \\

This paper focuses  on  the low-luminosity  galaxy population of  the  X-ray  bright  galaxy group  around  the   giant elliptical    NGC~5846      in
the     Local  Supercluster.     The    group   was    first       catalogued  in       \citet{devaucouleurs}.        Since  then, the          system
has   appeared   in    many       similar  catalogues   \citep{gellerhuchra,garcia}.  \citet{zabludoffmulchaey}   identified   13  dwarfs    in    the
central            region     of             the  system.      \citet{mahdavi},     hereafter       [MTT05]   state        $251\pm10$            group
members   including           83 spectroscopically        confirmed     ones.   \citet{carrasco}    have discovered  16   additional     low   surface
brightness    dwarfs. The NGC~5846   group  of galaxies    is  located   at   a   distance of    26.1 Mpc   in   the  Virgo III   Cloud of   galaxies,
a major   component of    the Local  Supercluster  \citep{tully82}.  Being   an elongated  structure   ranging from   the   Virgo  cluster   out    of
the    Supergalactic Plane,     this  cloud    contains   about   40     luminous   galaxies  with      the  NGC~5846    system     as   the   biggest
aggregate in   the   region. The   NGC~5846     group    is    the   most   massive   of    only     three  dense  groups  (NGC~4274, NGC~5846     and
M96)    in the Local Supercluster   that   are   dominated      by  elliptical  and    lenticular    galaxies,   being  the     third    most  massive
aggregate  of  early-type     galaxies   (after the     Virgo   and       Fornax clusters)        in      the       local     universe.      Moreover,
with           an  average          galaxy       density\footnote{Densities       taken      from       the      Nearby        Galaxies        Catalog
\citep{tullyfisher},     see \citet{tully88}   for  details.}  of $\rho=0.78\pm0.08$  Mpc$^{-3}$  compared        to $\rho=0.49\pm0.06$     Mpc$^{-3}$
for the  Local  Group       and $\rho\sim5$  Mpc$^{-3}$ for   the   central regions   of  the     Virgo  cluster, the    NGC~5846     system  presents
also  one    of   the  densest galaxy     environments  found  for     poor   groups.     The  group   is  dominated   by   early-type   galaxies with
a  massive  central elliptical  surrounded  by a   symmetric     X-ray halo.   The large   early-type  fraction,  the prevalence   of  dwarf  galaxies
together  with  the strong   and extendend   diffuse  X-ray     emission marks    the   NGC~5846 group   as an    evolved   system.   The    group  is
composed  of    two smaller  aggregates    around    the  two  brightest ellipticals       NGC~5846     and    NGC~5813. [MTT05]   determine       the
group´s    virial      mass  to       be     8.3         $\pm$   0.3  $\times$10$^{13}$        M$_{\odot}$         via  the  median      virial   mass
estimator      proposed     by  \citet{heisler}  and  derive a   velocity dispersion of  322    km s$^{-1}$.

The   goals  of this  work  are  to     study   new    faint  members  using  the  {\bf   S}loan   {\bf   D}igital  {\bf  S}ky   {\bf  S}urvey   (DR4)
\citep{adelman} to  extend  the  list  of  known  group members   and to   study  photometric  scaling relations, the distribution  of   morphological
types,   as  well  as the  spectroscopic   properties of the   low-luminosity  galaxy population.   The  paper is    organized as  follows. Section  2
gives    a   brief    overview    on  the     sample   selection   and    data   analysis procedures.    Section 3 focuses     on  the  low-luminosity
galaxy     population   and   presents   the    photometric  and   spectroscopic    data   of all    individual objects.  Finally,  the   results  are
summarized  and discussed  in Section     4. A  group distance of 26.1 Mpc    based  on  the  work of    [MTT05] is     used    throughout  this paper
corresponding to a distance  modulus  of $m-M=32.08$. The  systemic velocity   of the group is  assumed   to be represented    by NGC~5846    with   a
value of   $v_{r}=1710$    km    s$^{-1}$.      Concerning             the              separation             between  dwarf and  bright    galaxies,
an absolute  magnitude  of      $M_{B}=-16$    is    used \citep{fergusonbinggeli}   except   otherwise  stated. Total magnitudes  presented  in  this
work are   SDSS  model magnitudes \footnote{SDSS    model magnitudes  are  based   on a   matched galaxy  model  as optimal measure   of the flux of a
galaxy. The  code  fits    a  deVaucouleurs profile as     well as  an  exponential   profile to   the  two-dimensional  image. The best-fit model  is
stored as  the model magnitude.}.

\section {Sample Selection and Data  Analysis}
In order to extend   the   existing sample  of  low-luminosity  galaxies   in the  NGC~5846  system,  we   have queried the  SDSS    for  all galaxies
with $cz<3000$  km  s$^{-1}$   within  a   radius of  $2\degr=0.91$   Mpc,   similar   to     the second turnaround    radius  $r_{2t}=1.85$\degr   as
presented by [MTT05]\footnote{Bound  group members uncouple from  the   Hubble flow  at first  turnaround, i.e.   the zero-velocity  surface    around
the  group  \citep     {sandage},    begin to   collapse,  and    again   expand    to   a      radius of  second   turnaround,   finally  oscillating
and exchanging kinetic    energy with   other     group  members  \citep {bertschinger}.}.      Galaxies   beyond   these   limits   are  very  likely
to  be    characterised as field  or    background       objects.    The  query  resulted    in   a   total    sample of   74   galaxies:   19  bright
objects  and  55 dwarfs (see  Table~\ref  {faint} and Figure~\ref{chart})    with seven   objects found in     the outer   parts ($>1.33\degr$)     of
the  system  that show     no    entry in  previous  catalogues.  Though being   slightly   fainter    than  $M_{B}=-16$,  UGC 9751  and UGC 9760  are
listed  as bright members  due to their classification as Scd and Sd in NED\footnote{\tt  http://nedwww.ipac.caltech.edu/}.

Corrected   FITS   frames   (bias    subtracted,   flat-fielded   and   cleaned    of  bright   stars)   have    been   extracted   from   the    SDSS
for   these   55   dwarfs    in the $g^{\prime}$,  $r^{\prime}$ and   $i^{\prime}$ bands\footnote{See \citet {fukugita} for a description of the  SDSS
photometric system.}.   SDSS  model magnitudes have  been   adopted  from   the   database  for each  individual   galaxy  and transformed   to    the
Johnson-Morgan-Cousins    $B$   band       via    transformation     equations   from    \citet{smith}.    Absolute          magnitudes  were  derived
assuming an average    extinction  value of $A_{\rm{B}}=0.22$ from  \citet{schlegel} for   the whole   galaxy population. An  isophote    analysis was
carried out for    the  dE   subsample using    the  \texttt     {ellipse}      task  within    the      \texttt   {IRAF}   \texttt   {stsdas} package
to     fit  ellipses       to the  galaxy images      and   measure   the      deviations from   purely     elliptical  isophote shapes.   A  detailed
description    of   the procedure   is  given  by    \citet{jedrzejewski}.    The  analysis   resulted    in   surface  brightness profiles   and  the
harmonic content    of     the isophotes. A   \citet{sersic}   law   $\mu   (r)  =   \mu  _e  +  1.086~b_n~({\left(  {r/r_e }   \right)^{1/n}   - 1})$
with an   effective      radius  $r_{e}$, an effective     surface     brightness $\mu_{e}$   and a shape   parameter       $n$   as   free parameters
was   fit     to     each    object.   The quantity $b_{n}$   is   a   function   of  $n$   and  chosen  so  that   half     the galaxy     luminosity
is     located    within     the    effective  radius    \footnote{The  exact    relation     between  $n$     and    $b_{n}$    is   derived       by
$\Gamma(2n)=\gamma(2n,b_{n})$,   where  $\Gamma(a)$   and $\gamma(a,x)$    are the complete  and  incomplete gamma   functions, respectively. A   good
approximation    is $b_{n}=1.9908n-0.3118$ for $0.5\le  n \le 2$, which    has  been used  in  this work.}.    The     fitting was      carried    out
using          a   Levenberg-Marquardt    algorithm  until  a      minimum      in     $\chi^{2}$    was achieved,       yielding the      parameters
$n$,  $r_{e}$,       and       $\mu_{e}$   with      the corresponding   uncertainties.       The innermost   parts    of the   surface     brightness
profiles      ($\le1.4$       arcsec)     have   not  been  taken     into           account     for  the           fitting   procedure     due     to
seeing.  Central        surface            brightnesses $\mu_{0}\hspace{-1pt}_{_{B}}$    have  been  derived   from     the fit.       The     results
of     the       isophote    analysis         are presented      in    Table~\ref{surface}. S\'{e}rsic   models    have  been subtracted    from   the
original       images      yielding     residual frames    which are   investigated      for    photometric   substructures.  In addition  to     the
photometric   data,    wavelength    and      flux calibrated  spectra  (sky     subtracted  and     corrected   for  telluric absorption)  have  been
examined    for  all       sample  galaxies.     The     spectra cover    a    wavelength   range    of  $\lambda\lambda3800-9200$  \AA~and have  been
checked    for    spectral  lines    in  both   emission and    absorption.   Identified  spectral    lines     were     fit  by    Gaussians yielding
central  wavelengths     and   full      widths     at half maximum    (FWHM).    The  derived     central  wavelengths    were  subsequently  used to
verify SDSS redshifts.

\section {The Low-Luminosity Galaxy Population}

\begin{figure}
\centering
\includegraphics[width=\columnwidth]{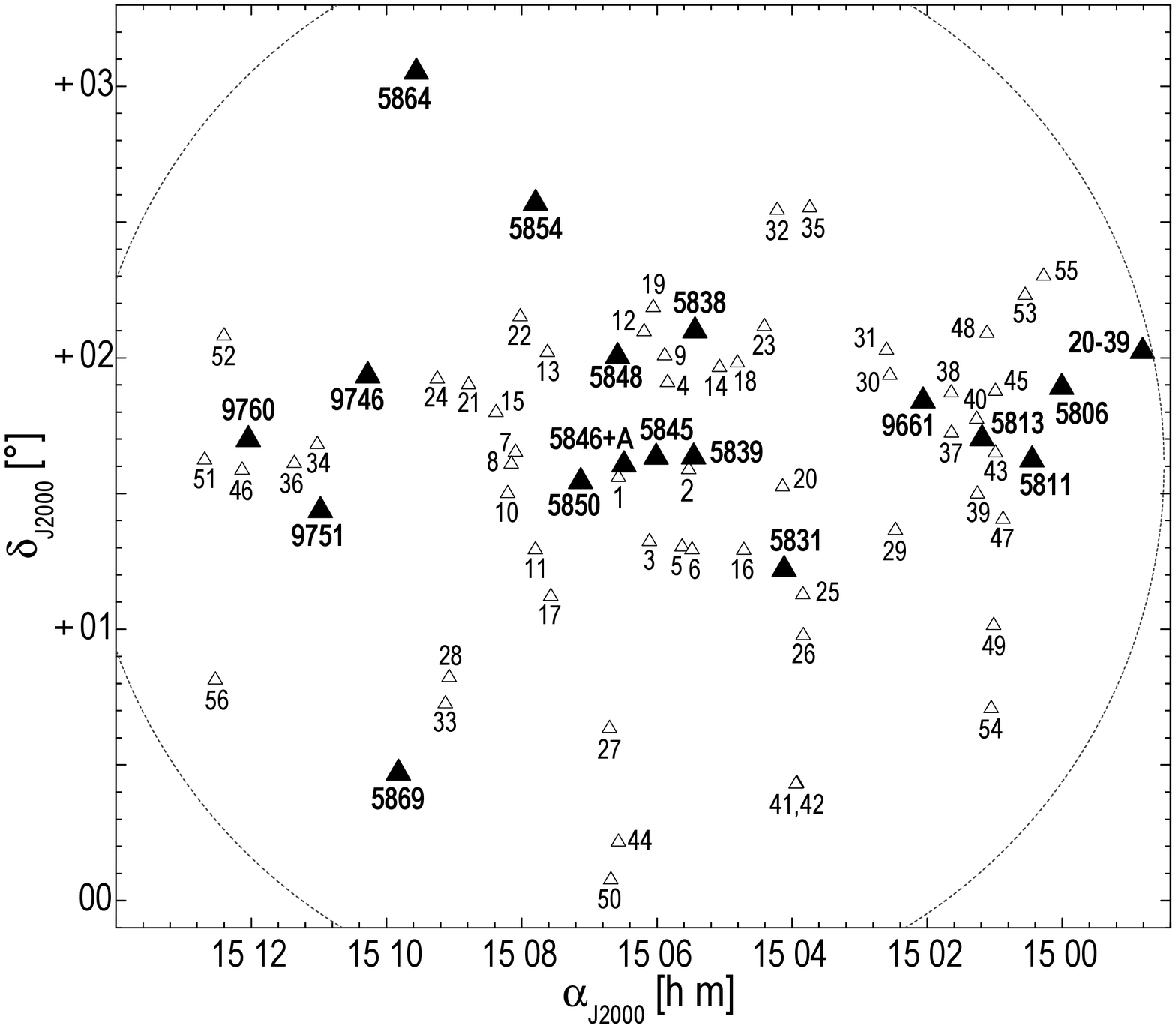}
\caption{\label{chart}Projected spatial   distribution  of    NGC~5846    group  members     as  studied   in this     work. Filled  triangles indicate
bright  members, while open  ones  represent    the low-luminosity  galaxies.   Numbers refer  to  the  galaxy identification   of  Table~\ref
{faint}. The circle shows the 2\degr (0.91 Mpc) query radius.}
\end{figure}

\begin{figure}
\centering
\includegraphics[width=\columnwidth]{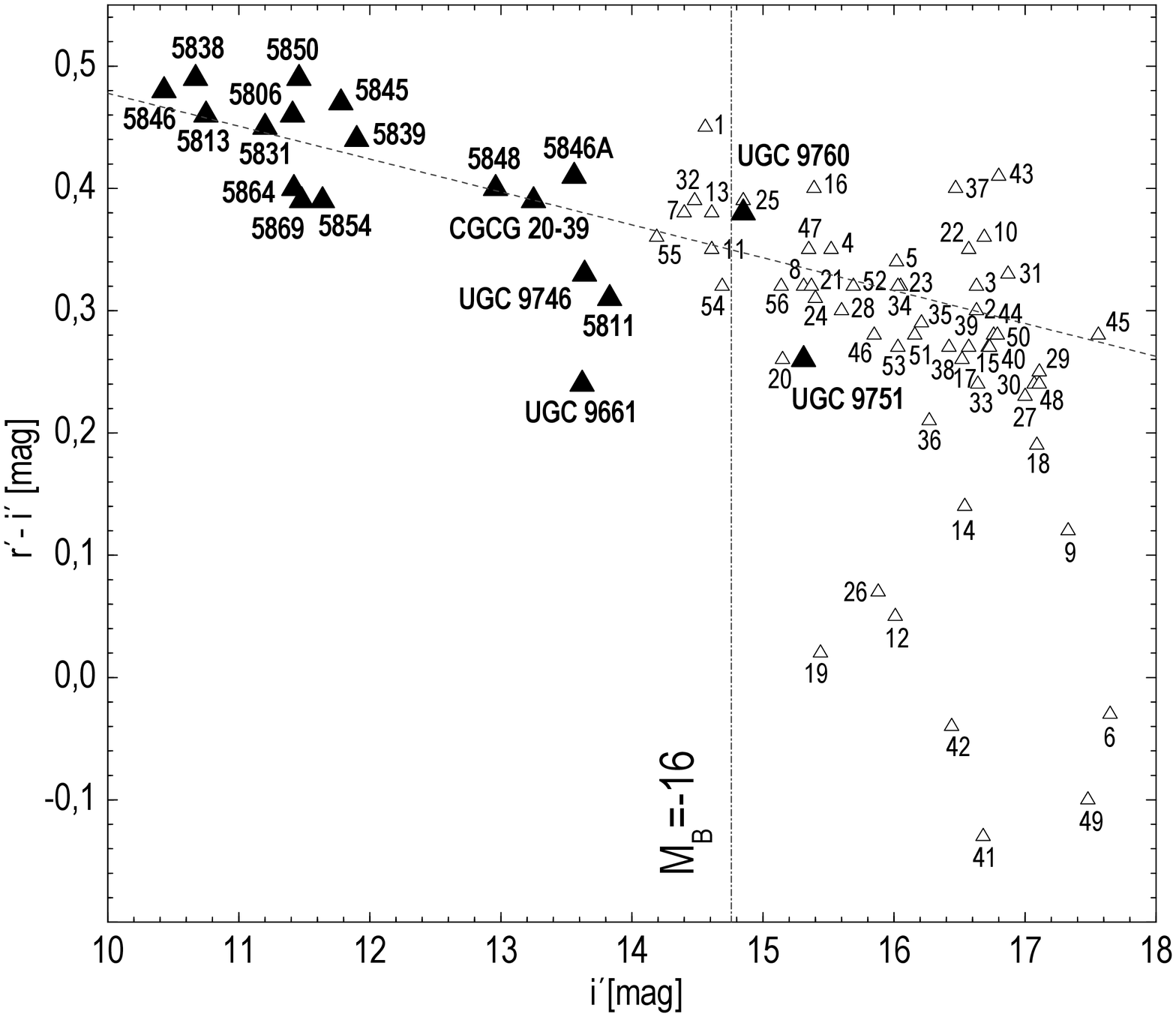}
\caption{\label{cmd}Colour-magnitude diagram of  NGC~5846   group    members. Numbers  as    in Figure~\ref   {chart}.  The    dashed  line  indicates
the red sequence   linear fit to   bright galaxies and  early-type  dwarfs: $r^{\prime}$$-$$i^{\prime}=0.75(\pm  0.05)-0.027(\pm0.003)$. The  vertical
line separates dwarfs from bright galaxies.}
\end{figure}

\begin{figure}
\centering
\includegraphics[width=\columnwidth]{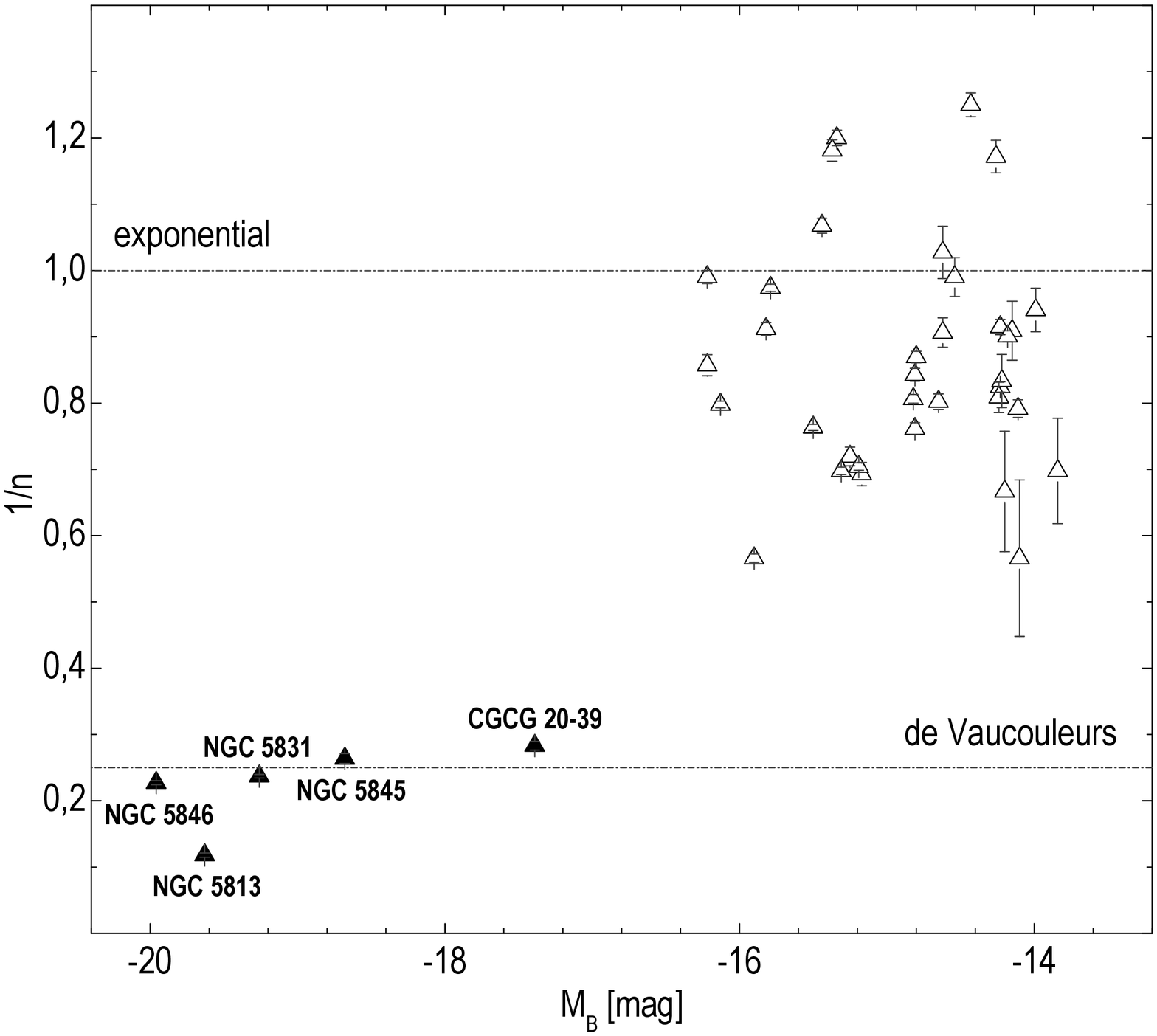}
\caption{\label {sersic}S\'{e}rsic shape   parameter  versus  absolute  blue   magnitude of   early-type  galaxies  in  the   NGC~5846   group.   Open
symbols   represent  dwarf  ellipticals  while   filled  triangles  show  bright  ellipticals.  The  horizontal  lines  indicate exponential  and de
Vaucouleurs r$^{1/4}$ laws.}
\end{figure}

Table~\ref   {faint}  lists    the  NGC~5846  galaxy   group   sample  as    studied   in        this work.   Morphological   types   for dwarfs  have
been  classified through visual  inspection   of  SDSS   images  and spectra. Objects  with  a  smooth   light  distribution  and exponential profiles
were   classified   as dEs.  The   dE ellipticity     was   determined     from  the      isophote  analysis. Dwarfs showing a blue patchy  appearance
and  emission lines   were classified  as dIrrs.      Morphological  types     for  bright  galaxies  have      been  taken   from NED.        Dwarfs
are        identified       with   increasing     projected     radial    distance  from NGC~5846. 

We  extend      previous group  member lists  with      seven    new   objects     in   the     outer    parts        ($>1.33\degr$)   of   the system
showing       no   entry         in previous catalogues.     Despite       exhibiting   slightly  brighter      values       than    $M_{B}=-16$, some
galaxies  have     been classified   as  dwarfs   due to       their    nearly     exponential   surface   brightness     profiles     or        their
compactness.   35   dwarfs  ($\sim$63\%)   have     been    identified            as         dEs          according                to            their
optical   appearance       and  red   colours  $(B-V)_{0}\simeq0.85$.     18   objects    ($\sim$32\%)    are      dwarf     irregulars      showing a
patchy,   blue    optical  appearance      with emission   lines  indicating      ongoing       star   formation. Three   galaxies  ($\sim$1\%) reveal
photometric fine    structure and  are in  the transition  to  bright     galaxies.           11   dwarfs    ($\sim$31\%  of        the      dEs)  are
found to be nucleated on basis of the analysis of the surface brightness profiles.

Our  list   of  nucleated  dwarfs  is   nearly  identical to  the  list  obtained by  [MTT05]  with   exception of  only   three   objects classified
differently. Taking      the        Schechter  luminosity    function derived by [MTT05],  we   conclude   that  with    an      absolute    magnitude
of   $M_{B}=  -13.32$     for  our faintest     object,     the  sample of  our work comprises  at  most 34\%    of  the   total number  of   galaxies
expected in   the  system down to  a limiting magnitude of  $M_{B}=-10$. Our sample yields  a dwarf-to-giant-ratio (DGR)  of  $\sim$3 as  lower limit.
Following  the photometric dwarf  classification of [MTT05],  this value could rise  to $\sim$12, confirming that the NGC 5846 group is indeed one  of
the richest groups in the Local Supercluster.

\addtocounter{table}{1}

The  projected   spatial   distribution   of     all    investigated   galaxies  is  illustrated     in   Figure~\ref  {chart}.   The     distribution
indicates     two   substructures     of    the  system around     the    two brightest    ellipticals   NGC~5846   and    NGC~5813.  A   third,
however    much  smaller    subgroup   can   be  found around UGC   9760,  which   in contrast    to the    NGC~5846  and NGC~5813    aggregates shows
no    noteworthy     X-ray      counterpart.  We present  on Figure~\ref    {cmd}    a   colour-magnitude     diagram   for      the    studied  group
galaxy     population.  The  diagram   indicates     the   prevalence  of   an       early-type dwarf morphology     and  the  continuation    of    a
red sequence similar   to    that  found  in    galaxy  clusters \citep{gladders}   into  the dwarf regime. Figure~\ref  {sersic} shows the  $n-M_{B}$
relation  consisting of  our dE sample  as  well as  the bright  galaxies classified  as   ellipticals   in    NED.   S\'{e}rsic  shape     parameters
of     the   low-luminosity    galaxy    population      show  a mean   value of  $n=1.19$  with  a scatter of  $\sigma_{n}=0.23$  representing nearly
exponential       surface     brightness    profiles      as      expected            for dwarf    ellipticals.   Three dwarfs (N5846\_10,  N5846\_15,
N5846\_43) show comparatively large error bars of the shape parameter due  to their faintness or contamination of the surface brightness profile by a
central nucleus.  Our   data    confirm the  trend  of   more  luminous        galaxies showing    greater  shape      parameters,  thus    a  steeper
surface  brightness     profile  than    the  less  luminous  objects.  Interestingly     there are   no   objects     with   intermediate  S\'{e}rsic
parameters   between   $2<n<3$.   NGC~5813 shows   a   comparatively high shape parameter  around $n\sim8$.

\subsection       {Scaling relations}
\begin            {figure*}
\centering
\includegraphics  [width=450pt]{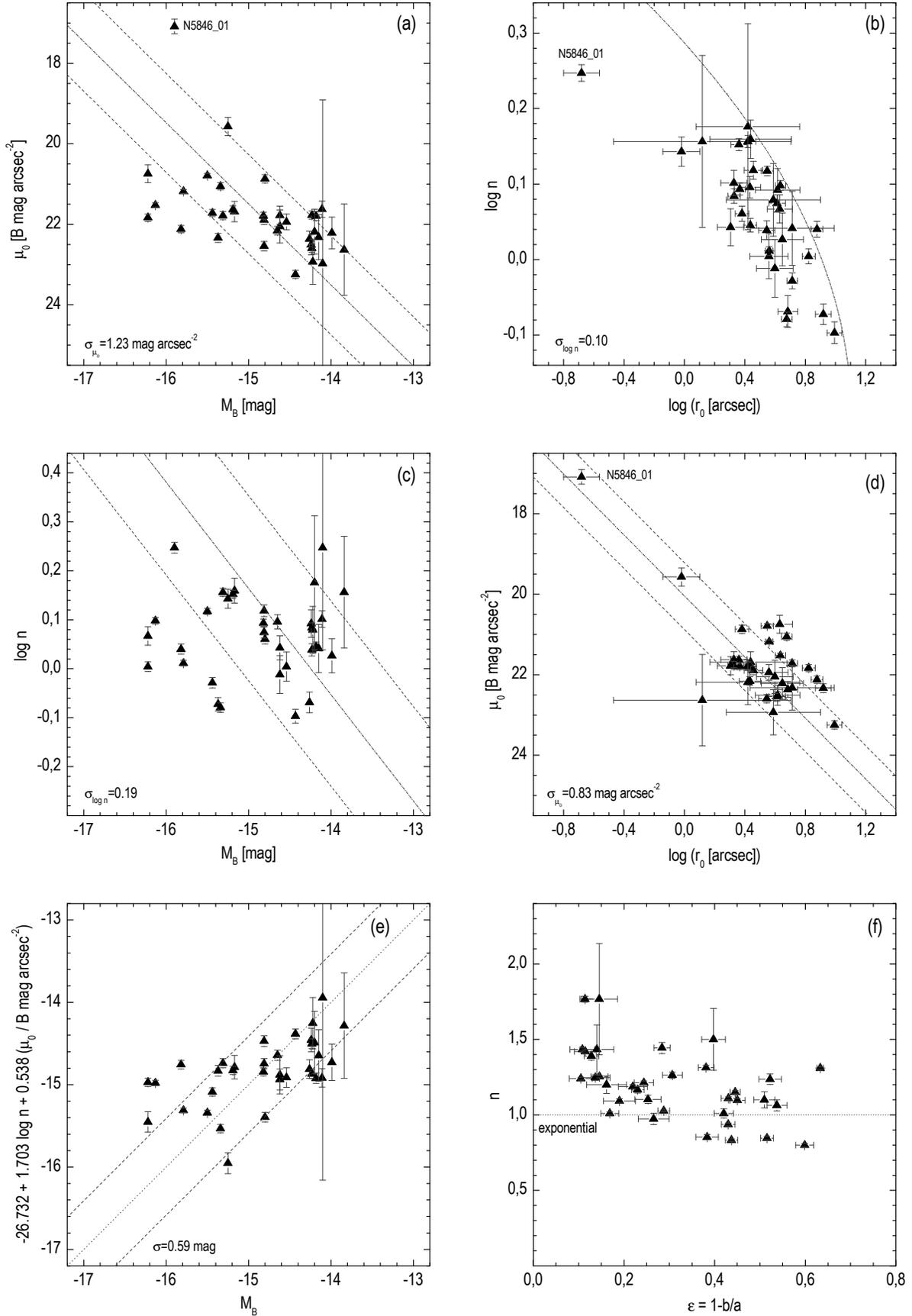}     
\caption          {\label  {scaling}  Photometric  scaling relations   of the   low-luminosity  galaxy  population in   the  NGC~5846 group.  (a)-(d):
dash-dotted   lines are  the   relations    of the Virgo   cluster   early-type     dwarfs  from [BJ98]  assuming  a   distance to  the  Virgo cluster
of  16.5 Mpc.  The  scatter of   our data  with  respect  to the   [BJ98] relations  is  given  in the   lower left  corner  of each panel.  $1\sigma$
deviations are    indicated as   dashed lines.   (a):  $\mu_{0}\hspace{-1pt}_{_{B}}-M_{B}$    plane. N5846\_01,  classified as  ultra-compact dwarf by
[MTT05] is  located  outside   $2\sigma$. (e):  best-fitting linear     combination    of    shape parameter    $n$ and   central   surface brightness
$\mu_{0}$  with  respect      to   absolute    magnitude   $M_{B}$.    N5846\_01    was  excluded    from   the     fit   and    is  not  shown in the
diagram. The dotted line  indicates  identity.  (f):   S\'{e}rsic  shape  parameter   versus  ellipticity  . Objects      with  high  ellipticity tend
to   have shallower surface    brightness profiles. The dotted  line indicates an exponential profile.} 
\end              {figure*}

When  addressing   the  question    of    the influence of the environment   on   the   evolution    of   galaxies,    scaling        relations     of
the photometric properties      of         early-type   galaxies     are      a   useful        tool.  However   the      environment   does  not seem
to affect  all early-type  systems   likewise.  The fundamental   plane   of  bright ellipticals  is  found  as     much  independent   of environment
as possible, while dwarf   ellipticals show a  stronger    variation  \citep{nieto,petersoncaldwell}. Elliptical galaxies and   bulges     are   known
to show     a   tight   correlation      in      the    $\mu-M_{B}$     plane   with      higher     surface    brightnesses    indicating     smaller
total luminosities    \citep{kormendy}.  Dwarf     galaxies    exhibit        an   opposite          trend,  being        a        distinct      class
of  objects     with      possibly  different   formation      and        evolution   histories.   \citet{binggelijerjen},  hereafter  [BJ98]     have
investigated      the    $\mu-B$  plane     for      Virgo   cluster   early-type     dwarfs      and      confirm   this  trend.          Figure~\ref
{scaling}(a)      shows         the  $\mu_{0}\hspace{-1pt}_{_{B}}-M_{B}$ plane       for our       dEs and     the    relationship   taken from   the
[BJ98](dash-dotted line) sample   of Virgo    dwarf ellipticals (assuming  a  Virgo distance  of 16.5   Mpc  corresponding  to a  distance modulus  of
$\sim$31.09).  Remaining relations  \ref     {scaling}(b)-(f)  focus   on        $M_{B}$,  $\log    n$,  $\mu_{0}\hspace{-1pt}_{_{B}}$,   $\log
r_{0}$  and     ellipticity $\epsilon$.     S\'{e}rsic     shape   parameters   used    here    should  not     be    mixed     up    with       those
of   [BJ98]: $n=1/n_{\rm{BJ98}}$.  The  parameter  $r_{0}$ is   the  scale    radius of     the S\'{e}rsic    model: $I(r)=I_{0}\exp[-r/r_{0}]^{1/n}$.
Table~\ref    {surface}  presents     the corresponding  values  of    the  isophote  analysis. Objects    with    larger  errors  than     the   data
itself   are not listed.   

Our data  in  the  $\mu_{0}\hspace{-1pt}_{_{B}}-M_{B}$ plane        are in  agreement   with  the work     of [BJ98].   One   object  falling  outside
the $2\sigma$   limit   is     N5846\_01   with   a  comparatively   too    high   surface  brightness   with    respect   to   its   luminosity  (see
section \ref{individual}).  In the   $\log n-M_{B}$   plane,   our  data show   a larger   scatter $\sigma_{\rm{log n}}=0.19$. Compared to  the  dwarf
sample  of the Virgo cluster, our dwarfs match  the Virgo dwarf  relation, however. Our sample also confirms the  trend  between shape  parameter  $n$
and  scale radius $r_{0}$ with smaller   objects exhibiting  higher shape  parameters. Again N5846\_01 is  the  only object in  our   sample  to  fall
off    the trend.  The     scale radius    $r_{0}$ shows    also  a    strong correlation     with the   central surface    brightness as    seen   in
the $\mu_{0}\hspace{-1pt}_{_{B}}-\log  r_{0}$  plane. The  NGC~5846  dwarfs  match with   the  relation obtained  for the Virgo   dwarfs with a  lower
scatter in magnitude   ($\sigma=0.83$mag  arcsec$^{-2}$)  than   [BJ98]  who  derive   a  value  of   $\sigma=1.25$mag  arcsec$^{-2}$.   Similarly  to
[BJ98]    we  also derived  the   best    fitting linear  combination  between     shape parameter  $n$, central   surface  brightness   $\mu_{0}$ and
absolute $B$   band magnitude with a  scatter of  $\sigma=0.59$mag. Finally,   we  also  checked for   a  correlation   between  shape   parameter $n$
and  ellipticity. There is a  slight trend  of  flattened   objects showing shallower surface brightness profiles.

\begin{figure}
\centering
\includegraphics[width=230pt]{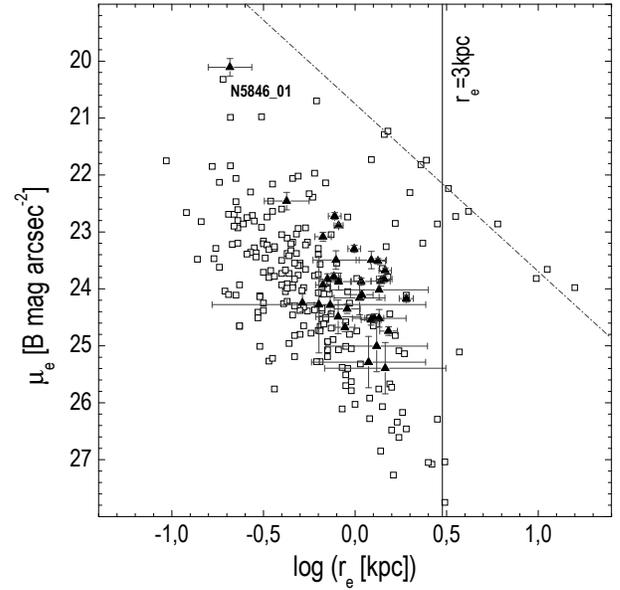}
\caption{\label{hamabe}$\mu_{e}-\log(r_{e})$   plane for   early-type   galaxies.   Effective  surface  brightnesses   are    shown in  the    Johnson
$B$  band.    The vertical   solid   line   separates  bright   and   regular    ellipticals      while   the   dash-dotted   line    represents   the
Hamabe-Kormendy relation   for elliptical  galaxies and bulges.  Black triangles   indicate  the  NGC~5846  faint  galaxy sample.  Open squares  refer
to the  groups studied by \citet{khosroshahi} for comparison.}
\end{figure}

The     structural    properties     of    early-type      systems    can     furthermore    be     investigated     in     the  \citet{hamabekormendy}
relation,   a photometric    projection  of     the fundamental     plane   of     galaxies relating     the  logarithm     of  effective radii    and
effective   surface   brightnesses  linearly  \citep{ziegler,diserego}.   This relation   divides    early-type    systems     in   \emph{regular} and
\emph{bright}   classes    separated     at   the effective   radius    of    $r_{e}\simeq  3$   kpc \citep{capaccioli}.   In  hierarchical  evolution
scenarios,  regular   galaxies  are   thought  to     be  the    progenitors   of    bright ones    evolving    through    succesive   mergers   along
the Hamabe-Kormendy    relation   \citep{capelato}. However,  recent    numerical   simulations  of   \citet{evstigneeva}  suggest   that    low  mass
systems like  dwarf galaxies    can  only     follow this  scenario  with     a  large   amout    of    dissipation  involved.   Figure~\ref  {hamabe}
shows  the Hamabe-Kormendy    relation  for  the   faint   galaxy     population       of       our    work.      Open      squares     are     dwarfs
from  X-ray  dim  and   X-ray     bright       groups   studied   by  \citet{khosroshahi}.  All   faint      galaxies  of    our    sample   represent
regular  systems   and  show  properties   similar to   the dwarfs  from  \citet{khosroshahi}   suggesting no  strong  difference  of the   structural
properties  of the  early-type dwarfs  from the NGC~5846   group compared  to  other group environments.  Again, N5846\_01 is the  only dwarf  clearly
separated from the main  cloud of dEs.

\begin{table*}
\begin{minipage}[t]{\textwidth}
\caption{\label{surface}Surface  photometry    of  the    low-luminosity  galaxy     sample.  Surface    brightnesses  are    shown  in     magnitudes
arcsec$^{-2}$, effective radii  in arcseconds.  Numbers  in parantheses indicate errors  of the   last significant digit.  The quality of the fits  is
shown in  the last three columns.}
\centering
\renewcommand{\footnoterule}{}
\begin {tabular}{lcccccccccc}
\hline\hline
\multicolumn {1}{c}{galaxy}   &  $n_{i^{\prime}}$ &   $n_{r^{\prime}}$ &   $n_{g^{\prime}}$  &  $\mu_{0}\hspace{-1pt}_{_{B}}$           &   $r_{e_{i^{\prime}}}$  &  $r_{e_{r^{\prime}}}$ &  $r_{e_{g^{\prime}}}$  &  $\chi_{\nu,i^{\prime}}^{2}$  &  $\chi_{\nu,r^{\prime}}^{2}$  &  $\chi_{\nu,g^{\prime}}^{2}$    \\
\hline                                                                                                                                                                                                                                                                                                                     
N5846\_01                     &      1.84(4)      &      1.82(3)       &       1.64(3)       &                 17.1(2)                  &         1.6(4)          &        1.7(3)         &         1.6(3)         &             0.00043             &             0.00025             &              0.00055              \\
N5846\_02                     &      1.12(3)      &      1.06(2)       &       1.10(2)       &                 22.6(1)                  &         6.9(8)          &        6.9(6)         &         7.0(5)         &             0.00528             &             0.00317             &              0.00229              \\
N5846\_03                     &      1.08(1)      &      1.14(2)       &       1.11(2)       &                 21.8(1)                  &         5.1(2)          &        5.7(4)         &         5.9(5)         &             0.0018              &             0.00325             &              0.00221              \\
N5846\_04                     &      1.43(2)      &      1.53(2)       &       1.30(2)       &                 21.63(9)                 &         8.4(9)          &        9.5(9)         &         7.6(5)         &             0.00159             &             0.00064             &              0.00099              \\
N5846\_05                     &      1.31(2)      &      1.27(1)       &       1.14(2)       &                 21.79(9)                 &         6.1(5)          &        6.3(3)         &         5.7(3)         &             0.00288             &             0.00109             &              0.00217              \\
N5846\_09                     &      0.91(9)      &      1.09(9)       &       1.3(1)        &                 22.3(6)                  &          9(3)           &         10(4)         &         12(7)          &             0.00681             &             0.00957             &              0.0078               \\
N5846\_10                     &      1.4(3)       &       2.0(7)       &       1.9(8)        &                  23(4)                   &            -            &           -           &           -            &             0.00468             &             0.00335             &              0.00513              \\
N5846\_11                     &      1.01(3)      &      1.15(3)       &       1.34(5)       &                 20.7(2)                  &         9.0(9)          &         10(1)         &         10(2)          &             0.00824             &             0.00925             &              0.01286              \\
N5846\_13                     &      1.22(2)      &      1.33(1)       &       1.21(1)       &                 21.52(6)                 &         11.3(8)         &        12.3(5)        &        10.9(7)         &             0.00286             &             0.00065             &              0.00066              \\
N5846\_15                     &      1.8(6)       &      1.20(8)       &       1.5(1)        &                 22.2(6)                  &            -            &         10(3)         &         13(7)          &             0.0038              &             0.00293             &              0.00142              \\
N5846\_16                     &      1.42(2)      &      1.45(2)       &       1.43(2)       &                 21.8(1)                  &         10.0(9)         &        10.1(8)        &         10(1)          &              0.001              &             0.00069             &              0.00106              \\
N5846\_17                     &      0.97(6)      &      1.11(6)       &       0.95(3)       &                 21.9(2)                  &          6(1)           &         7(2)          &         5.6(7)         &             0.00375             &             0.00238             &              0.00143              \\
N5846\_21                     &      1.37(2)      &      1.21(1)       &       1.35(1)       &                 20.79(6)                 &          11(1)          &        9.7(5)         &        10.8(6)         &             0.00332             &             0.00085             &              0.0015               \\
N5846\_22                     &      1.23(2)      &      1.23(2)       &       1.18(2)       &                 21.8(1)                  &         5.2(4)          &        5.3(3)         &         5.3(4)         &             0.00354             &             0.00212             &              0.00356              \\
N5846\_23                     &      1.41(3)      &      1.31(3)       &       1.22(2)       &                 21.9(1)                  &          9(1)           &         8(1)          &         8.7(7)         &             0.00163             &             0.00327             &              0.00156              \\
N5846\_24                     &      0.98(2)      &      0.93(1)       &       0.90(2)       &                 21.72(9)                 &         8.0(5)          &        7.7(4)         &         7.7(5)         &             0.00107             &             0.00118             &              0.0026               \\
N5846\_25                     &      1.09(1)      &      1.02(1)       &       0.97(1)       &                 21.18(5)                 &         6.5(3)          &        6.3(3)         &         6.4(2)         &             0.00508             &             0.00374             &              0.0028               \\
N5846\_27                     &      0.91(3)      &      0.81(2)       &       0.84(4)       &                 22.4(2)                  &         6.9(7)          &        6.0(5)         &         6.3(9)         &             0.00099             &             0.00075             &              0.0024               \\
N5846\_28                     &      0.80(1)      &      0.81(1)       &       0.89(2)       &                 21.05(9)                 &         6.5(4)          &        6.0(3)         &         5.8(4)         &             0.00345             &             0.00289             &              0.00406              \\
N5846\_31                     &      1.02(6)      &      1.10(5)       &       1.07(8)       &                 22.2(4)                  &          8(2)           &         9(2)          &          8(2)          &             0.00173             &             0.0018              &              0.00227              \\
N5846\_32                     &      0.99(2)      &      0.99(1)       &       1.05(2)       &                 21.83(9)                 &         10.9(9)         &        11.3(7)        &        11.9(9)         &             0.0016              &             0.00131             &              0.00168              \\
N5846\_33                     &      0.78(2)      &      0.79(2)       &       0.83(2)       &                 23.3(1)                  &          11(1)          &         12(1)         &         13(1)          &             0.00362             &             0.00173             &              0.00329              \\
N5846\_34                     &      1.20(2)      &      1.23(3)       &       1.13(2)       &                 22.5(1)                  &         9.5(8)          &         10(1)         &         9.4(9)         &             0.00161             &             0.00255             &              0.00225              \\
N5846\_35                     &      1.28(4)      &      1.26(3)       &       1.20(2)       &                 22.2(1)                  &          7(1)           &        7.1(9)         &         7.3(7)         &             0.00541             &             0.00358             &              0.00206              \\
N5846\_37                     &      1.3(1)       &       1.1(1)       &       1.2(1)        &                 22.9(6)                  &          11(6)          &         7(2)          &         10(6)          &             0.00111             &             0.00376             &              0.00245              \\
N5846\_38                     &      1.08(5)      &      1.02(4)       &       1.21(5)       &                 21.8(2)                  &         4.0(6)          &        3.8(5)         &         4.4(7)         &             0.00575             &             0.00515             &              0.00399              \\
N5846\_39                     &      0.74(6)      &      0.99(3)       &       1.19(9)       &                 22.1(4)                  &          6(1)           &        6.1(6)         &          7(2)          &             0.0024              &             0.0019              &              0.00173              \\
N5846\_40                     &      1.15(7)      &      1.38(5)       &       1.18(6)       &                 22.5(3)                  &          10(3)          &         13(3)         &          9(2)          &             0.00661             &             0.0016              &              0.00416              \\
N5846\_43                     &      1.3(2)       &       1.8(4)       &       1.2(2)        &                  23(1)                   &          6(5)           &           -           &          4(3)          &             0.00621             &             0.00222             &              0.00715              \\
N5846\_46                     &      1.36(4)      &      1.36(5)       &       1.45(5)       &                 19.6(2)                  &         3.4(6)          &        3.3(7)         &         3.3(8)         &             0.0148              &             0.0148              &              0.00764              \\
N5846\_47                     &      0.87(2)      &      0.82(2)       &       0.85(2)       &                 22.3(1)                  &          11(1)          &        10.6(9)        &         11(1)          &             0.00337             &             0.00217             &              0.00371              \\
N5846\_50                     &      1.17(4)      &      1.27(3)       &       1.35(4)       &                 21.6(2)                  &         5.5(8)          &        5.8(8)         &          6(1)          &             0.00782             &             0.00516             &              0.00784              \\
N5846\_51                     &      1.05(2)      &      1.21(2)       &       1.19(2)       &                 20.9(1)                  &         4.9(3)          &        5.5(5)         &         5.4(4)         &             0.00177             &             0.0034              &              0.00283              \\
N5846\_52                     &      1.14(4)      &      1.78(9)       &       1.41(5)       &                 21.7(3)                  &          8(1)           &         14(8)         &         10(3)          &             0.00397             &             0.00245             &              0.00322              \\
N5846\_56                     &      1.10(2)      &      1.09(2)       &       1.10(2)       &                 22.12(9)                 &          15(1)          &         15(1)         &         15(1)          &             0.00268             &             0.00264             &              0.00186              \\
\hline
\end{tabular}
\end{minipage}
\end{table*}

\begin{table*}
\begin{minipage}[t]{\textwidth}
\caption{\label{emission}Spectral properties of emission-line dwarfs. Fluxes present the integrals of the fitted  Gaussians and are given in units  of
10$^{-17}$ ergs s$^{-1}$ cm$^{-2}$. H$\alpha$ and H$\beta$  fluxes have  been corrected for extinction. Numbers in parentheses indicate errors  of the
last  significant digit.}
\centering
\begin{tabular}{lcccccccccc}
\hline\hline
                                                      &    H$\beta$     & $[$O{\scriptsize~III}$]$ & $[$O{\scriptsize~III}$]$ & $[$N{\scriptsize~II}$]$ &    H$\alpha$    & $[$N{\scriptsize~II}$]$ & $[$S{\scriptsize~II}$]$ &  $[$S{\scriptsize~II}$]$  &           SFR           &  $12+\log$O/H  \\
\multicolumn{1}{c}{\raisebox{1.5ex}[-1.5ex]{galaxy}}  & $\lambda$4861.3 &     $\lambda$4958.9      &     $\lambda$5006.8      &     $\lambda$6548.1     & $\lambda$6562.8 &     $\lambda$6583.4     &     $\lambda$6717.0     &      $\lambda$6731.3      &  M$_{~\odot}$yr$^{-1}$  &     [dex]      \\
\hline
N5846\_06                                             &      348.1      &           45.8           &          128.4           &          14.0           &      992.0      &          51.7           &          66.6           &           47.7            &        0.0064(3)        &     8.2(1)     \\
N5846\_07                                             &      619.6      &            -             &           4.2            &          62.5           &     1765.9      &          173.7          &          107.2          &           81.3            &        0.0114(4)        &     8.4(1)     \\
N5846\_08                                             &        -        &            -             &            -             &           7.7           &      51.0       &           7.2           &          12.5           &            8.0            &            -            &     8.5(1)     \\
N5846\_09                                             &      79.0       &            -             &           18.9           &           3.1           &      225.1      &           8.5           &          21.4           &           16.5            &            -            &     8.1(1)     \\
N5846\_12                                             &      300.1      &           93.9           &          270.4           &           7.9           &      855.4      &          16.6           &          47.5           &           26.7            &        0.0055(3)        &     7.8(2)     \\
N5846\_14                                             &      409.9      &           71.9           &          212.0           &          10.1           &     1168.3      &          47.8           &          89.3           &           64.0            &        0.0075(4)        &     8.1(2)     \\
N5846\_15                                             &        -        &            -             &            -             &            -            &      24.2       &            -            &            -            &             -             &            -            &       -        \\
N5846\_17                                             &        -        &            -             &            -             &            -            &      46.7       &           9.1           &           6.5           &            6.8            &            -            &     8.6(1)     \\
N5846\_18                                             &        -        &            -             &            -             &            -            &      30.7       &            -            &           7.7           &            7.4            &            -            &       -        \\
N5846\_19                                             &      43.7       &            -             &           19.3           &           8.6           &      124.6      &          14.0           &          21.3           &           12.3            &            -            &     8.4(1)     \\
N5846\_20                                             &      171.1      &           5.4            &           39.9           &          17.5           &      487.5      &          42.4           &          51.4           &           37.5            &            -            &     8.4(1)     \\
N5846\_27                                             &        -        &            -             &            -             &            -            &      38.8       &            -            &           7.9           &             -             &            -            &       -        \\
N5846\_28                                             &      298.8      &           15.6           &           50.1           &          21.2           &      851.5      &          67.1           &          64.5           &           52.6            &        0.0055(3)        &     8.3(1)     \\
N5846\_33                                             &        -        &            -             &            -             &            -            &      29.7       &            -            &            -            &             -             &            -            &       -        \\
N5846\_38                                             &      103.8      &           12.1           &           40.8           &           8.6           &      295.8      &          20.5           &          16.2           &           14.4            &            -            &     8.3(1)     \\
N5846\_39                                             &      34.0       &            -             &           10.4           &            -            &      96.8       &           9.3           &          19.2           &           12.7            &            -            &     8.4(1)     \\
N5846\_40                                             &        -        &            -             &            -             &            -            &      42.8       &            -            &           9.3           &            5.9            &            -            &       -        \\
N5846\_41                                             &     2576.2      &          1222.0          &          3664.0          &          20.6           &     7342.2      &          68.1           &          182.0          &           127.8           &        0.0473(9)        &     7.6(2)     \\
N5846\_42                                             &      914.9      &          475.5           &          1385.0          &          10.9           &     2607.5      &          31.4           &          76.4           &           48.9            &        0.0168(5)        &     7.7(2)     \\
N5846\_46                                             &      223.7      &           15.3           &           43.8           &          16.6           &      637.6      &          51.7           &          53.9           &           43.1            &        0.0041(3)        &     8.3(1)     \\
N5846\_49                                             &      28.9       &            -             &           8.7            &            -            &      82.3       &           2.4           &           5.3           &            7.7            &            -            &     8.0(1)     \\
N5846\_50                                             &        -        &            -             &            -             &            -            &      32.2       &            -            &            -            &             -             &            -            &       -        \\
N5846\_51                                             &      62.6       &           8.9            &           38.8           &           8.6           &      178.4      &          15.8           &          24.5           &           20.9            &            -            &     8.4(1)     \\
N5846\_54                                             &      290.9      &            -             &           45.4           &          31.4           &      829.2      &          89.8           &          75.4           &           57.6            &        0.0053(3)        &     8.4(1)     \\
N5846\_55                                             &      461.1      &            -             &           17.9           &          47.0           &     1314.2      &          140.1          &          81.2           &           61.9            &        0.0085(4)        &     8.4(1)     \\
N5846\_56                                             &      40.6       &            -             &           15.6           &          12.0           &      115.7      &          10.5           &          14.1           &           10.2            &            -            &     8.3(1)     \\
\hline
\end{tabular}
\end{minipage}
\end{table*}

\subsection       {\label{individual}Comments on individual objects}
\begin            {figure*}
\centering
\includegraphics  [width=420pt]{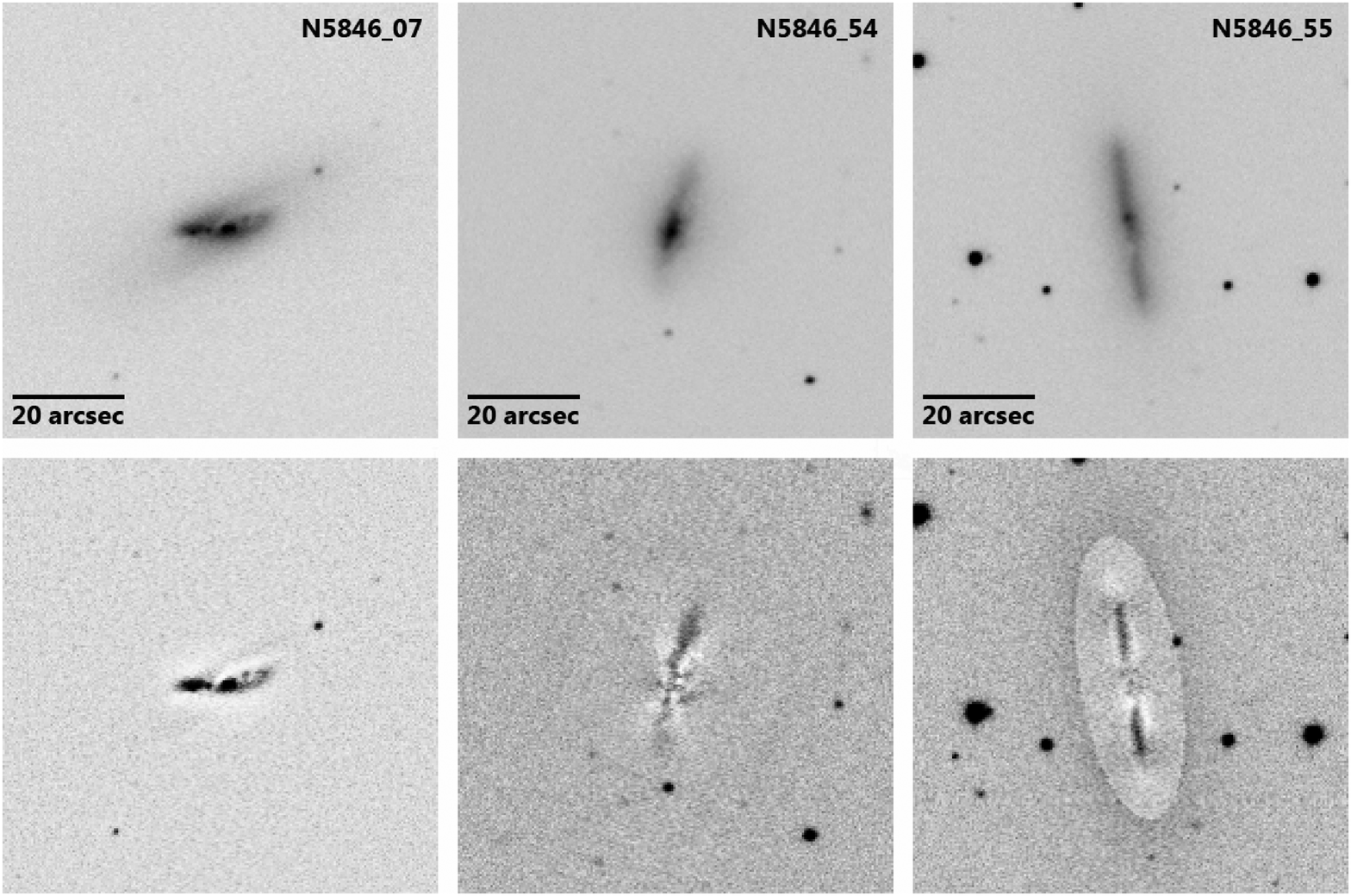}     
\caption          {\label   {fine} Modeling   of   faint   galaxies with     significant fine     structure. Figures    show SDSS    $r^{\prime}$ band
images (upper    panel)  and     residual  images     (lower panel) of N5846\_07, N5846\_54  and N5846\_55.  Since  no S\'{e}rsic model  was  obtained
for N5846\_07, gaussian smoothing  was applied to the galaxy.} 
\end              {figure*}

Most of  the  investigated  dwarfs do not  show any  photometric    or  spectroscopic peculiarities. A    few galaxies  in  the   sample  are    worth
a   closer  look,   however. Figure~\ref {fine}   presents  all    objects  which evidently exhibit  fine structure in  their optical appearance  with
$r^{\prime}$ band images and  residual  frames  shown  for each  galaxy.  These objects  lie  furthermore in the   transition   to bright galaxies due
to   their relatively large angular diameter.\\

{\bf N5846\_01}  is the  closest dwarf elliptical to NGC~5846  with a   much higher   central  surface brightness than the  rest   of our dwarfs.   It
is furthermore the most reddish galaxy  from our  low-luminosity  galaxy sample.  With    an average   effective  radius   of   only  $r_{e}\sim$  200
pc  in   all   studied passbands,  this     object   is also the   most   compact     of    our  sample.   Due    to   the     visual  appearance  and
the     close proximity    to NGC~5846,    [MTT05] suggested   that N5846\_01  has very     likely   been  affected  by    tidal  stripping   and  can
therefore rather    be classified   as ultra  compact   dwarf.  This   idea     is also       supported by     the   fact    that the    nearby object
NGC~5846A shows  a similar   morphological appearance.  In fact, NGC~5846A is   one of the   extremely  rare   "compact   ellipticals" like M32,   the
prototypical  object   of  this    class. Thus,  NGC5846\_01   fits  perfectly into   the picture   that  the    structural differences    of  compact
ellipticals and    ordinary dEs     result   from   the fact   that  the compacts  formed  within  the potential  well    of another  massive   galaxy
whereas  dEs  evolved  as  isolated   systems \citep{burkert}. 

{\bf N5846\_07} is  classified   as S0/a  galaxy by  [MTT05]   and  originally considered    as almost certainly  background  object in  their  sample
due to its   optical appearance. Spectroscopy confirmed  the group membership  however. Assuming this distance, the   galaxy has a projected  diameter
of  $D_{25}\sim7$ kpc which    locates   the  galaxy  in  the   transition between  bright and    dwarf galaxies. The spectrum   of  N5846\_07 reveals
strong H$\alpha$ emission  together with the  weaker H$\beta$, [NII]   and  [SII] features    indicating ongoing star  formation.   Two spiral    arms
are  identified  in   the   optical  image which  could    also be interpreted as  signatures of ongoing interaction  with the nearby irregular  dwarf
N5846\_08,  only 21kpc away.

{\bf  N5846\_41/42}  are    two   large  HII    regions   of    the    irregular  dwarf  KKR15     \citep{huchtmeier, karachentseva}    classified  as
individual  galaxies in  SDSS.  The host  galaxy shows an  extremely blue  colour,  the  majority of  light contained in  the two   HII  regions  with
the brighter  one, N5846\_41,  being the  bluest object  of our   sample.  Due   to the high concentration  of light, N5846\_41/42 present  by far the
highest S/N ratio of  all spectra  in   our sample. Besides  the usual H$\alpha$,   H$\beta$,  [OIII], [NII]  and [SII] features  both  spectra   also
show    the  weaker  Balmer lines   H$\gamma$,     H$\delta$,  H$\xi$    along  with    the    more   uncommon   HeI     $\lambda$5875.6,  [Ne    III]
$\lambda$3868.7   and  ArIII $\lambda$7135.8  transitions. Of    all emission-line  galaxies  in   our sample,   N5846\_41/42 show  the  lowest oxygen
abundance   with an average value of 12+log(O/H)=7.68 (see section \ref{emissiondwarfs}).

{\bf N5846\_54} and  {\bf N5846\_55}  were   both classified as  S0  galaxies in    the [MTT05] sample   and are the   only objects  of  our work that
show photometric substructures. With    a projected linear  extent  of  $D_{25}\sim4$ and  7  kpc  these objects    can,  similarly  to  N5846\_07, be
placed in  the  transition  to dwarf   galaxies.  The  surface brightness  profile  of N5846\_55 exhibits a  rather  flat gradient  up  to  14 arcsec,
wherefrom   the  light    distribution  falls    much  steeper,   indicating    a  possible  bar   feature  \citep{gadotti}.  Spectroscopically, both
galaxies show ongoing  star formation.

\begin{figure}
\centering
\includegraphics[width=\columnwidth]{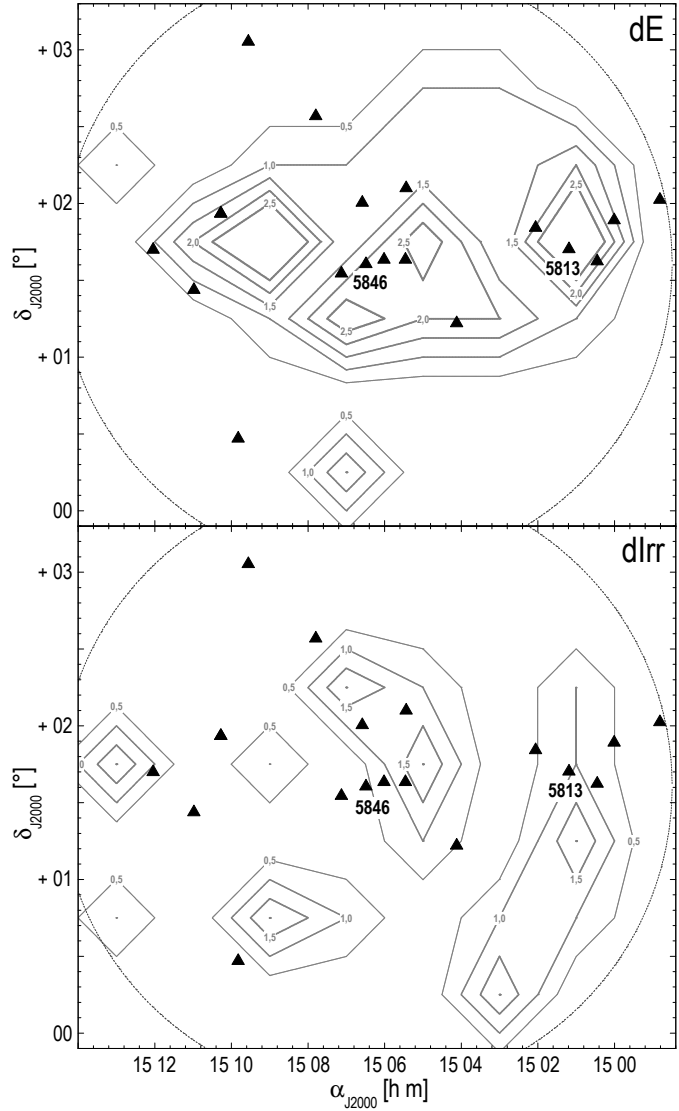}
\caption{\label{clustering}Clustering  properties for  both  early  and late-type  dwarfs of    the NGC~5846 system.  Contours  are   linearly spaced
and   show    the projected     number    density  of   the    distinct morphological   populations.   Triangles    refer  to  bright  group  members.
The values give the number of galaxies  per 0.25 square degree.}
\end{figure}

\begin{figure}
\centering
\includegraphics[width=\columnwidth]{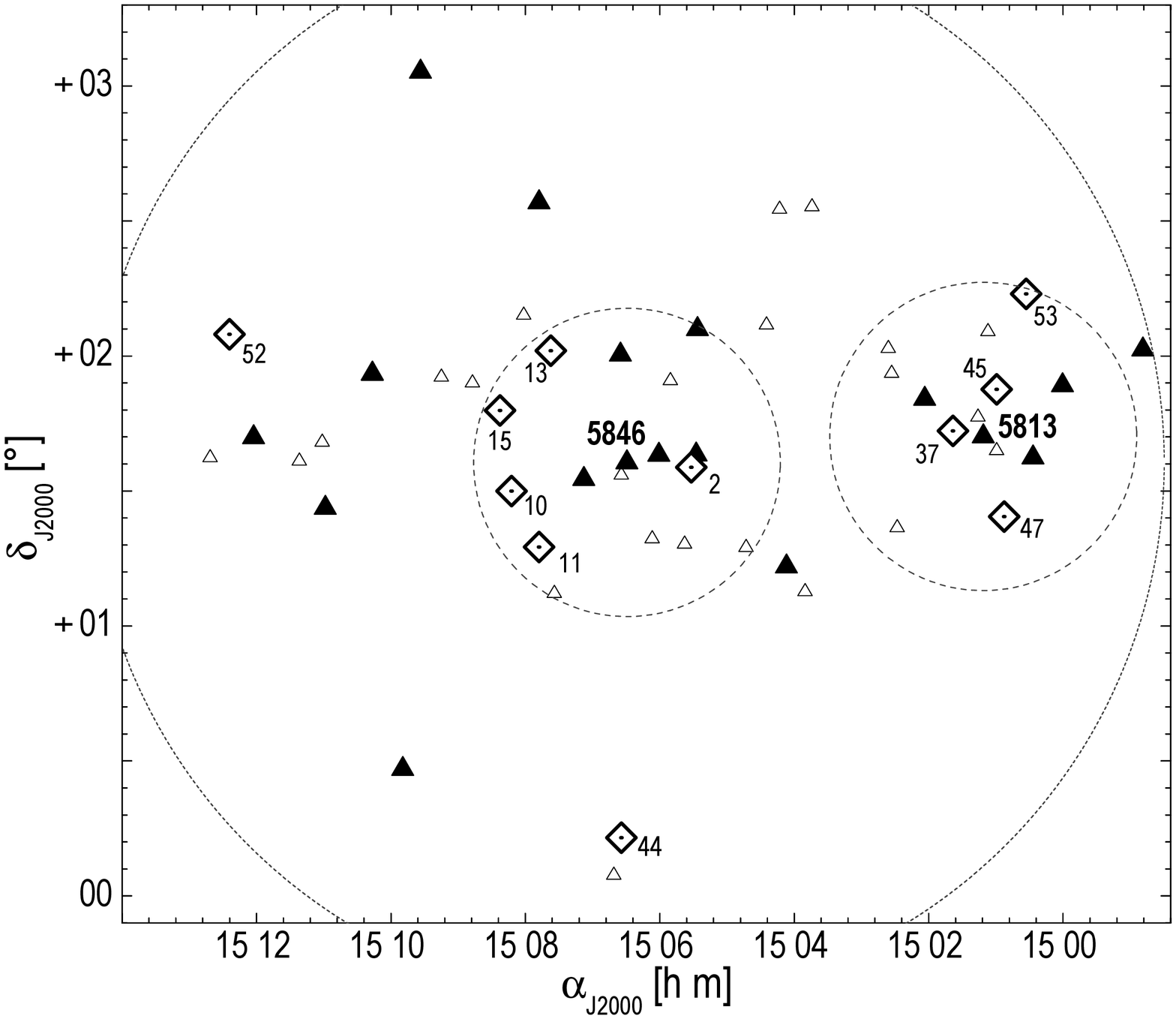}
\caption{\label{nuclei}Distribution of nucleated dwarfs  in  the  NGC~5846   group. Black triangles  are   bright group  members, open  triangles  dEs
without nucleus. Diamonds show nucleated  dwarfs, concentrated around NGC~5846 and NGC~5813 (circles indicate a radius of 260kpc).}
\end{figure}

\subsection{Spatial distribution of morphological types}
We have studied  the morphology distribution  for the  low-luminosity  galaxy population in  the NGC~5846  system  for both  the early- and  late-type
populations. Figure~\ref {clustering}   illustrates the     clustering for     both     the  dE     and  dIrr  samples    with   the projected  number
density   shown   separately   for    each    morphological  type.   The   diagram    displays   the       same  field   of   view   as   presented in
Figure~\ref   {chart}.   Density   contours   have      been   created     using    a        grid   with a    spatial     resolution   of     0.5\degr
$\times$ 0.5\degr. The  segregation between   early- and   late-type low-luminosity  glaxies   in  the NGC~5846 system  is  evident: dwarf  ellipticals
are found predominantly in  the vicinity of bright   galaxies while irregular   systems  are clearly  detached  from  this distribution  and diffusely
arranged  in the  outer regions of the  group.  Morphology  shows also a  strong  correlation  with the projected number  density. Early-type   dwarfs
populate  the high-density  regions    around  NGC~5846   and  NGC~5813,    while   irregulars show   comparatively   low  agglomeration.    Moreover,
the   satellite morphologies  resemble the  morphology  of their   elliptical hosts  (NGC~5846 and   NGC~5813).  Similar  results   have  been   found
by \citet{weinmann} who,  on the   basis of   a  large sample  of  SDSS  groups,   inferred that  the  satellite  morphologies in   groups   correlate
strongly with the   morphology of  the central   host galaxy.   Regarding the   radial  morphology  distributions,  Figure~\ref{clustering}  indicates
that   the projected  number    density  of  the  early-type  population    shows    a    much  higher    radial   gradient    than   the    irregular
subsample,  merely presenting  any   radial  variation.  In addition  to  the  optical picture, the diffuse X-ray  component correlates also  with the
early-type dwarf distribution  (see [MTT05]), indicating   the  dark matter  potentials of the  group.  We have   also checked  the  distribution   of
nucleated  dwarfs  within   the  NGC~5846  system    shown  in   Figure~\ref{nuclei}. With the exceptions  of two objects, all nucleated   dwarfs  are
found    in  the vicinity ($\sim$ 260kpc)  of NGC~5846  and  NGC~5813. Research on  the   distribution of   nucleated dwarfs  in    the  Virgo cluster
has also  shown that   nuclei  are predominantly found  in the  central regions    and not  in the  outskirts   \citep{binggelicameron}. \citet   {oh}
argued   that  this  could   be  explained    assuming   nuclei   to   be   the result     of orbital   decay   of   globular  clusters.    Through  a
series   of  numerical simulations    they   showed   that  for   dwarf   galaxies   exposed    to  little   external   tidal  perturbation  dynamical
friction  can   lead   to significant orbital  decays  of globular clusters   and  the formation   of  compact nuclei  within a   Hubble-time.   Thus,
a   larger    fraction   of nucleated   dwarfs   is   expected   in    the   centres   of   galaxy    clusters   where   the   extragalactic   tidal
perturbation  tends to preserve  the integrity of  dwarf galaxies unlike in the outskirts where this perturbation tends to be disruptive.

\subsection{\label{emissiondwarfs}Emission-line dwarfs}
Line                fluxes              from          H$\alpha$,           H$\beta$,              $[$N\begin{scriptsize}         II\end{scriptsize}$]$
$\lambda\lambda$6548.1,                 6583.4                                     \AA,                                         $[$O\begin{scriptsize}
III\end{scriptsize}$]$  $\lambda\lambda$4958.9,   5006.8    \AA,               and           $[$S\begin{scriptsize}              II\end{scriptsize}$]$
$\lambda\lambda$6717.0, 6731.3 \AA~features   together    with     star formation   rates       and    oxygen     abundances  have  been measured  for
dwarfs showing  emission  lines  in their  spectra. The  data   are     presented   in     Table~\ref     {emission}. Star formation    rates   (SFRs)
were   derived    via   H$\alpha$ fluxes    using     the      \citet{kennicutt98} relation:   $\textrm {SFR}(M_{\odot}  \textrm      {\space  year$^{
-1}$})=7.9      \cdot  10^{      -42}L(\textrm     {H}\alpha)(\textrm       {ergs       s$^{-1}$)}$.      H$\alpha$  extinction        was       taken
into  account        using  an    average       extinction           value            of     $A(\textrm{H}\alpha)=0.95$         mag    based        on
\citet{kennicutt83}          and  \citet{niklas}.  Additionally,      galactic    extinction      $A_{R}=0.15$     mag     was       also   taken into
account  so that      a      total   extinction value    of   1.10   mag     for  H$\alpha$         was  allowed.    Errors    for    star   formation
rates  were estimated assuming    H$\alpha$      photon  noise as    the   dominant    error   source.   Star     formation   rates      of   galaxies
exhibiting   H$\alpha$       fluxes     with        S/N$<$15   are  not            listed          in          Table~\ref{emission}.         Emission
characteristics    have         been     analysed                                                     calculating                                  the
line                                                                                                                                              flux
ratios                                                                          $\textrm{[N\begin{tiny}II\end{tiny}]}~\lambda6583/\textrm{H$\alpha$}$,
$\textrm{[O\begin{tiny}III\end{tiny}]}~\lambda5007/\textrm{H$\beta$}$,                                                                             and
$\textrm{[S\begin{tiny}II\end{tiny}]}~(\lambda6716+\lambda6731)/\textrm{H$\alpha$}$                            according                            to
\citet{veilleuxosterbrock}.   Oxygen   abundances        have                been                                         estimated              using
the    $N_{2}$      index              ($\textrm{[N\begin{tiny}II\end{tiny}]}~\lambda6583/\textrm{H$\alpha$}$)                 method               as
defined  by \citet{denicolo}:  $12+\log(\textrm{O/H})=9.12(\pm   0.05)+0.73(\pm    0.10)\times    N_{2}$.    Extinction   and    reddening     effects
together  with   specific    absorption   components   underlying      H$\alpha$    and   H$\beta$   can       considerably   affect       line   flux
measurements. A  correction was  applied    to     the   emission-line      galaxies    by      taking      H$\alpha$    extinction into account   and
assuming    a   Balmer decrement    of H$\alpha$/H$\beta=2.85$    for    HII  region-like galaxies \citep{veilleuxosterbrock}  to  consider   H$\beta$
extincion too.   

\begin            {figure*}
\centering
\includegraphics  [width=450pt]{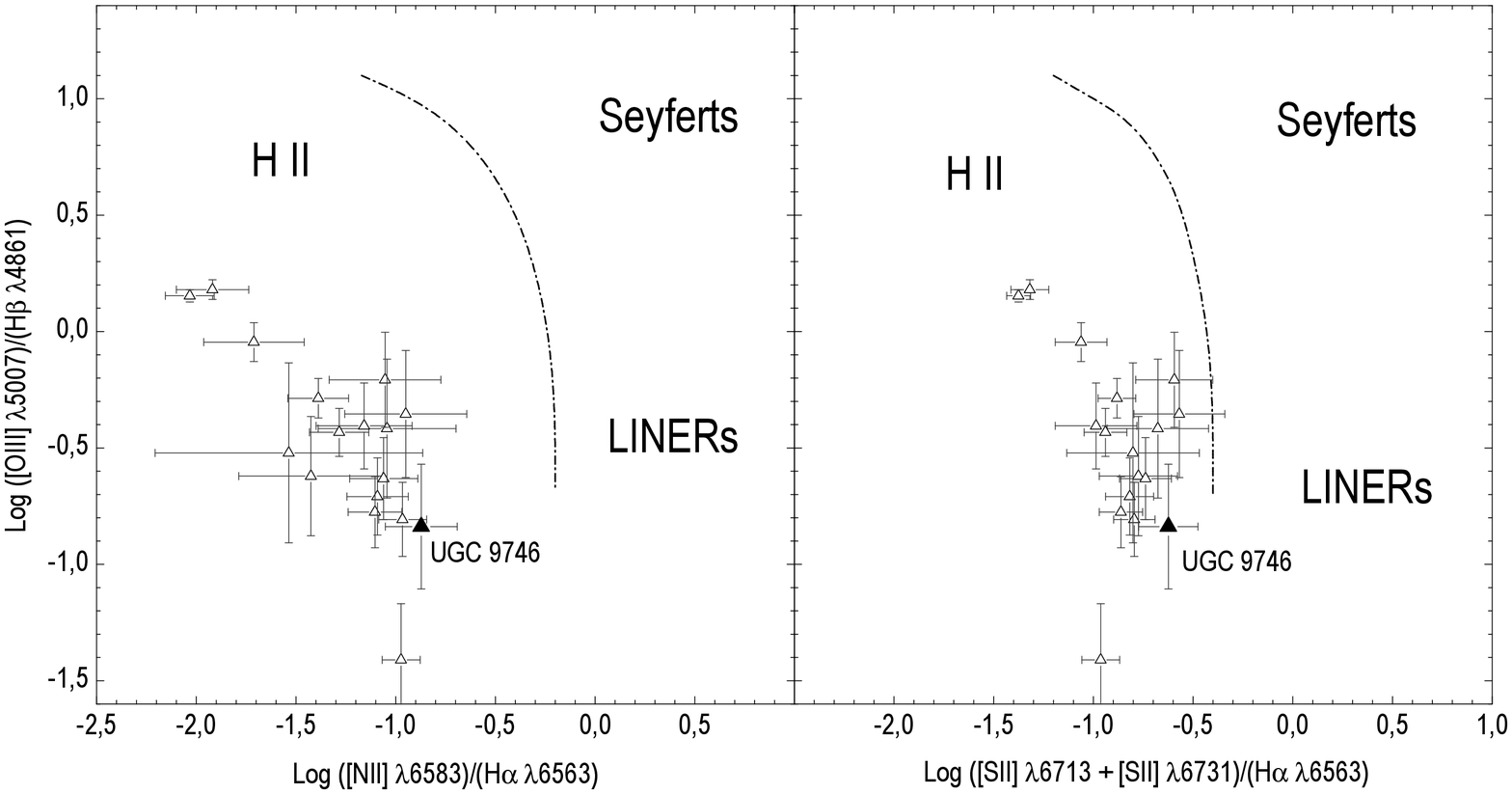}     
\caption          {\label{osterbrock}Classification of emission-line galaxies   in the NGC~5846  group   according to Veilleux \& Osterbrock   (1987).
The line   separates HII  region-like galaxies from AGNs. All measured objects show flux ratios explained by simple photoionization.} 
\end              {figure*}  

\begin            {figure}
\centering
\includegraphics  [width=\columnwidth]{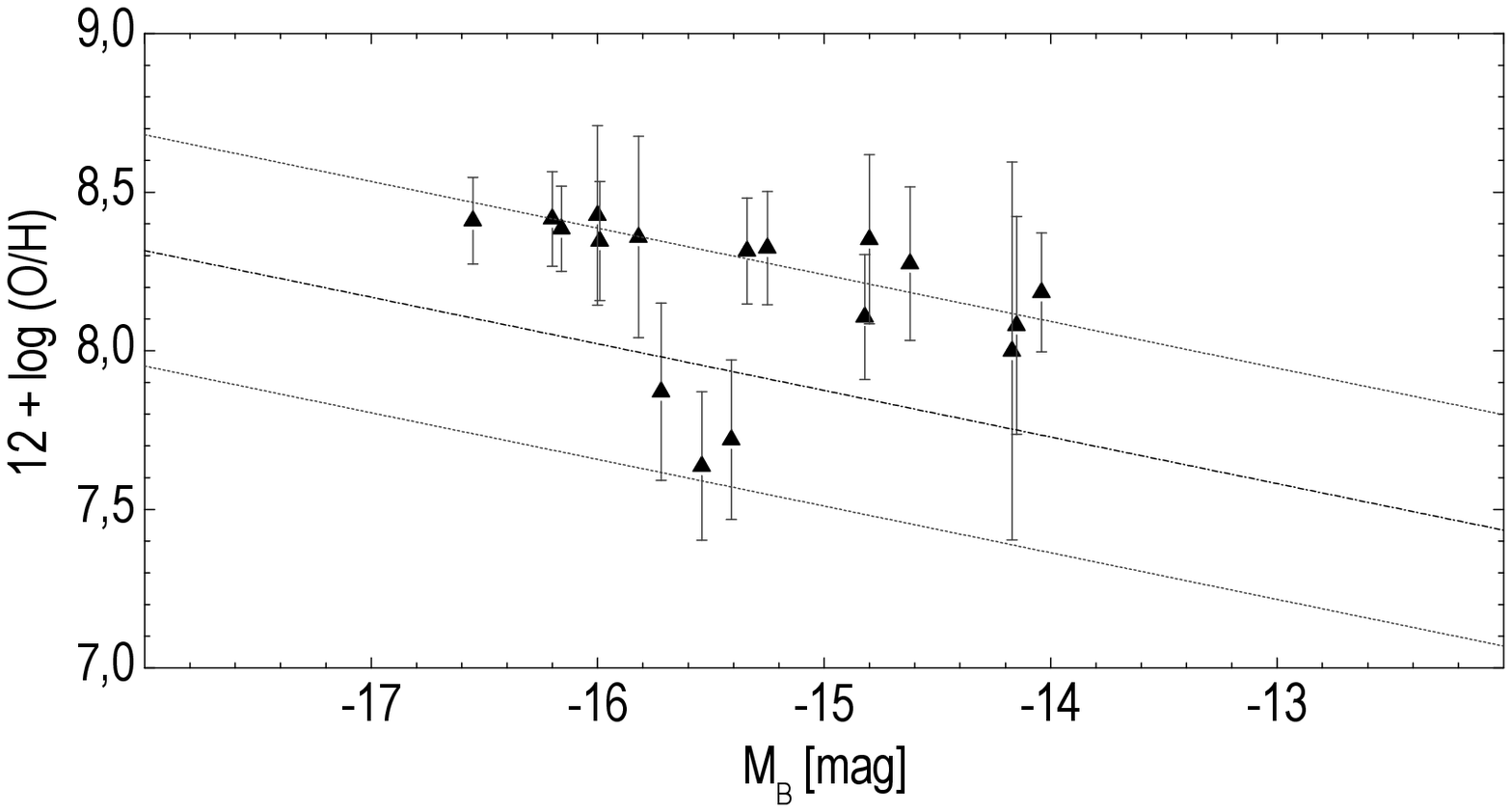}     
\caption          {\label{oxygen}Oxygen  abundances versus  abolute B  magnitudes of  NGC~5846  irrgeluar dwarfs.  The abundances have been determined
indirectly via   the $N_{2}$  method as  proposed by  Denicólo  et  al. (2002).  The dash-dotted  line shows the metallicity-luminosity  relation from
\citet{richer}. 1$\sigma$ deviations are shown as dashed lines.} 
\end              {figure}  

Figure \ref{osterbrock} distinguishes  HII  region-like galaxies from  AGNs  such  as  Seyferts  or    LINERS, according to the  diagnosis proposed by
\citet{veilleuxosterbrock}.  All  emission-line     galaxies  can    be explained     by          photoionization.  Only few  bright galaxies     have
SDSS spectra.  Furthermore   their  H$\beta$  flux  is   strongly  contaminated by  the   old  stellar population,  which makes   a proper measurement
of the  line ratio problematic.  In addition,   the  [OIII]  feature at   $\lambda5007$\AA~   is    not    seen    in    most  of     the investigated
bright              galaxies.        However,                just              from                the              measurements              of   the
$\textrm{[N\begin{tiny}II\end{tiny}]}~\lambda6583/\textrm{H$\alpha$}$                                                                              and
$\textrm{[S\begin{tiny}II\end{tiny}]}~(\lambda6716+\lambda6731)/\textrm{H$\alpha$}$  ratios,   all   these   objects  can    be  classified   as   HII
region-like  objects. There  is no   evident correlation   between  star    formation activity  and location    of emission-line  dwarfs  within   the
group. \citet{lequeux}   suggested  that     the  oxygen    abundances   correlate    with  total     galaxy  mass    for   irregular    galaxies with
more massive galaxies  exhibiting  higher metal   content.  Since  the  galaxy   mass  is    a  poorly  known  parameter,   the metallicity-luminosity
relation instead   of the  mass-metallicity  relation is   usually considered.   Figure~\ref{oxygen}   shows the   oxygen abundance  of  emission-line
dwarfs measured  via the  $N_{2}$ method versus absolute  blue   magnitude.  Most  of  our   dwarfs lie   above the   metallicity-luminosity  relation
from \citet{richer}. Only the   two  HII  regions N5846\_41/42    of  KKR15   \citep{huchtmeier} (see  Section  \ref{individual}) and  N5846\_12  fall
below this  sequence. The error    bars reflect  merely  the  uncertainty     of  the   $N_{2}$  flux    ratio   and   the  error   of    the   linear
least squares   fit   of \citet{denicolo}.

\section{Discussion and Conclusions}
We have undertaken an   analysis  of the   photometric  and  spectroscopic   properties  of  the low-luminosity  galaxy    population  of the NGC~5846
group  of galaxies     located   in   the  Virgo III   Cloud of    galaxies in    the Local  Supercluster.  The    group    is of particular  interest
since it is   the   most   massive   of    only     three  dense  groups  (NGC~4274, NGC~5846     and    M96)    in the Local Supercluster  that   are
dominated     by  elliptical and    lenticular    galaxies,  being  the    third    most  massive    aggregate  of early-type     galaxies  (after the
Virgo   and      Fornax clusters)       in      the       local      universe. Of the  19 bright galaxies in  our sample, E and S0 galaxies amount  to
$\sim58$\%. Seven  new   group   members in   the  outer   regions   ($>$80   arcmin)  of    the group    have   been   identified    via   SDSS   and
NED redshifts  complementing    existing     member     catalogues. \\

The  dwarf-to-giant-ratio  (DGR) of the   NGC~5846 system  is large:  our   sample of spectroscopically determined group  members  yields a  DGR    of
$\sim$3.  Taking into   account   the photometric classification   of   group members down to  $M_{R}=-10$   by [MTT05] this   value could rise up  to
$\sim$12. \citet{fergusonsandage} have  shown that  the  DGR increases with  the  richness of  a  group and   constitute an  early-type-dwarf-to-giant-ratio (EDGR),  (dE+dE,N+dS0 over  E+S0) of   5.77 for  the  Virgo  cluster  and  3.83 for the Fornax  cluster down   to $M_{B}=-13.5$. For  the  Coma
cluster \citet{seckerharris} state  an EDGR of $5.80 \pm 1.33$  down to $M_{B}=-13.5$. Our data yield  an  EDGR of 2.69 for the   NGC~5846 system down
to our faintest dwarf  ($M_{B}=-13.32$). This value  shows  that the NGC~5846  system has less early-type dwarfs per giant  early-type galaxy compared
to  clusters but   still a   much  larger  EDGR than   other  groups  (Leo:  1.48,  Dorado: 1.44,  NGC~1400:  2.08; all   down  to  $M_{B}=-13.5$, see
\citet{fergusonsandage}),   indicating  that  NGC~5846  is  indeed  one   of  the  more  massive aggregates in    the Local  Supercluster. \\

A    colour-magnitude diagram    of   the     investigated   group    members    reveals      a    red sequence displaying   the     well-known  trend
of   early-type    galaxies  exhibiting    redder       colours     (higher    metallicities)     for     higher luminosities    \citep{caldwell}.  In
contrast, irregular   dwarfs  do not  form a   sequence  but   show a   large spread  in   colours.   There is  no evident segregation between  bright
galaxies and dwarfs. \\

Photometric    scaling relations  have  been  studied   for  early-type dwarfs  and  compared with   the  relations   derived  by [BJ98]    for  Virgo
cluster    dwarf    ellipicals.   If      bright    ellipticals       and      early-type     dwarfs share    a    similar   evolution, a    continous
sequence   in    scaling  relations  with   respect    to    the   galaxy     luminosity    would  be    expected.  It   is      known that   in the
surface-brightness-luminosity    ($\mu-M$)       plane   \citep{kormendy},           early-type   dwarfs      and     bright   ellipticals    follow
different   trends,  marking dwarf         galaxies a   distinct    class   of objects      that have      undergone   a   different evolutionary path
compared to   ordinary ellipticals.    Our    data    match  the     trends  from   Virgo      cluster   dEs suggesting     a  similar  structure  and
origin    of     the dwarfs  in both      the    Virgo  cluster     and  the   NGC~5846   group. One  special  object    falling    off the       main
cloud   of  early-type  dwarfs is   NGC5846\_01,  however.  It   is   the  closest   dwarf     to  NGC~5846 and shows  a comparatively   high  surface
brightness and  compactness  with respect  to   the  other  dwarfs    investigated.   In  the       $\mu_{0}\hspace{  -1pt}_{_{B}}-M_{B}$   plane  the
object   lies        outside   the  $2\sigma$   limit       compared  to       the   relation      of [BJ98].        Due to     the proximity       to
NGC~5846,    the    galaxy  has      very likely        been   affected     by   tidal stripping   and   can  therefore  rather be  classified      as
ultra      compact      dwarf. This   idea is also     supported by    the   fact   that the    nearby object  NGC~5846A,   the  elliptical  companion
to NGC 5846,  shows   a similar morphological appearance,    though being   too  bright   to   be   considered as     dwarf. In  fact,  NGC~5846A   is
one  of  the    extremely   rare     "compact  ellipticals"  like     M32,  the      prototype    of     this   class.    NGC5846\_01   confirms   the
idea    that  the   structural differences   of  compact ellipticals and     ordinary dEs    result    from   the fact    that  the  compacts   formed
within  the   potential well   of another    massive galaxy whereas   dEs   evolved  as   isolated    systems \citep{burkert}.    Our  data also  show
that early-type     dwarfs  with   high  ellipticity tend      to  have    shallower   surface     brightness  profiles.   We    also    checked   the
location  of  the  NGC~5846  low-luminosity   galaxy population     in   the  $\mu_{e} -\log(r_{e})$      plane  with   respect to        the Hamabe-Kormendy   relation,    a     photometric projection     of the      fundamental   plane    of    galaxies        relating     the          logarithm
of  effective    radii          and    effective  surface  brightnesses      linearly.  The      NGC~5846   dwarfs   represent    regular     galaxies
in     the  classification        of    \citet{capaccioli}  indicating   that    these    systems   are   the     building      blocks     of     more
massive    galaxies,      in  agreement     with  $\Lambda$CDM  cosmology.  Moreover,     the   dwarfs      are  comparable      with  dwarfs     from
X-ray   dim         and   X-ray  bright  groups  studied   by \citet{khosroshahi}. \\

Scaling relations  for dwarf    ellipticals are  not  the  only  way    to enlighten  their  origin  and evolution. Morphological  segregation  within
clusters and    groups   also   gives   the   opportunity    to     test   the   formation  and    environmental  dependence      of    the      faint
galaxy population.   The   well-known      morphology-density     relation      \citep{dressler}  is     not   only   restricted     to    the  normal
Hubble types    but    is  also    found  to      apply    to     dwarf      galaxies  \citep{binggeli}.    This   relation  has  been investigated in
detail  in our    immediate vicinity, the Local  Group \citep{grebel}.   We    confirm      this morphology-density         relation     for     dwarf
galaxies in  the   NGC~5846   system.    While      early-type        dwarfs  are   concentrated        around  the         two   massive  ellipticals
NGC~5846 and    NGC~5813, objects     of          irregular    type  are       more    randomly   distributed       and  found  predominantly       in
the outer  regions.  This morphological   segregation  could    be  explained    by  gas    stripping   due   to   the    infall  of the    dwarfs  in
the group    centre.   This   scenario     was    also   investigated        by  \citet     {weinmann}  and    \citet{park},    who    inferred   that
satellite morphologies  in  groups  tend to    be     similar     to   those    of hosts.    With     the   comparatively     high   density   of  the
NGC~5846  system   and       the       hot   and       dense        halo    gas       present  around       NGC~5846     and         NGC~5813,     the
dependence of the  satellite  morphology   on host  morphology   can be      interpreted     by the hydrodynamic  and   radiative    interaction    of
the  hot  X-ray gas of   the  two host    galaxies with their   surrounding  satellites.  We have  also   checked  the   distribution   of   nucleated
dwarfs  in   the NGC~5846   system with    respect   to  non-nucleated  ones.    Interestingly, nearly   all    dwarfs    that     show    a   nucleus
superimposed      on the    underlying   smooth   surface  brightness   profile    are  found    in   the  vicinity   of  NGC~5846    and     NGC~5813
within a  radius    of $\sim$260kpc.   This       is     in agreement     with      the   work          of \citet{binggelicameron}      on  the  Virgo
cluster, who   show  that  dwarfs  located    near the  centre  of   the   cluster   are   mostly   nucleated   while those  in the    outskirts   are
non-nucleated.  Assuming    the formation      of   the    nuclei   due     to the     orbital  decay      of globular   clusters, \citet{oh}   argued
that this  could  be   explained    by  extragalactic  tidal  perturbation  which   tends to  preserve   the  integrity  of dwarf  galaxies at cluster
centres but tends to disrupt dwarf galaxies in their outskirts. \\

Only two objects  of  the low-luminosity  galaxy population show   photometric fine structure.  These galaxies exhibit   ongoing star  formation   and
diameters in        the   transition  to     bright  galaxies.  This    lack of  photometric  peculiarities  in the low-luminosity  galaxy  population
emphasises  the fact  that the NGC~5846  system is  an  old,   well-evolved aggregate where no   recent   interactions between  individual     members
have  occured.    Emission characteristics      and star    formation rates     of   the     irregular         dwarfs        show typical  values  for
 low-luminosity  galaxies,   additionally  supporting   the  idea   of  a    system   without   recent   activity   enhancing   star   formation.
Oxygen  abundances    have    been    derived    indirectly       via         the         $N_{2}$          index         method     proposed        by
\citet{denicolo}.   Our abundances    show a   large  scatter     ($\sigma=0.36$ dex) with    respect to    the  metallicity-luminosity    relation  of
\citet{richer} due    to the  intrinsical  scatter in  the  $N_{2}$   method  itself.   Nevertheless,   all  objects  outside   1$\sigma$   deviations
lie above   the  metallicity-luminosity  relation,  indicating a comparatively  high oxygen abundance for irregular dwarfs in the NGC~5846  group. \\

The study  of       the  low-luminosity   galaxy      population   of  the       NGC~5846  group    has revealed  that   the  dwarf galaxies  in  this
massive group  match  the   trends observed  for  dwarf  galaxies  in other aggregates   indicating a similar  formation  and evolution  scenario. The
spatial distribution of morphological types suggests that the structural properties of dwarf galaxies are clearly dependent on the location within the
group. The structural properties of the dwarfs confirm the evolved state of the system. The group  remains   special    due to      the     dominating
early-type   morphology,   the   strong      X-ray emission   and  the    large dwarf-to-giant-ratio compared to other groups. 

\begin{acknowledgements}
We acknowledge the useful comments from the anonymous referee which helped to improve the paper. PE has been supported  by  the University of   Vienna
in  the frame  of the  Initiativkolleg  (IK)  The Cosmic Matter  Circuit   I033-N. This work    has  made  use   of       the   astronomical      data
reduction    software      {\texttt     {IRAF}}    which    is    distributed     by  the  National   Opical  Astronomy Observatories,     which   are
operated   by   the   Association   of  Universities    for  Research   in   Astronomy Inc.,  under  cooperative agreement  with  the National Science
Foundation.
\end{acknowledgements}

\longtab{1}{                                                                                                                                                                                              
\begin{longtable}{lccccccccl}                                                                                                                                                                             
\caption{\label{faint}Galaxies within 2\degr around NGC~5846 as studied in this work.}\\                                                                                                                  
\hline\hline                                                                                                                                                                                              
\multicolumn {1}{c}{galaxy}  &  [MTT05]$^{a}$ & $\alpha_{2000}$$^{b}$ & $\delta_{2000}$$^{b}$ & $d$ [arcmin]$^{c}$ & $cz$ [km s$^{-1}$]  &  $i^{\prime}$$^{b}$  &  $r^{\prime}$-$i^{\prime}$$^{b}$  &  $M _{B}$  &  \multicolumn {1}{c}{type}   \\
\hline                                                                                                                                                                                                    
\endfirsthead                                                                                                                                                                                             
\caption{continued.}\\                                                                                                                                                                                    
\hline\hline                                                                                                                                                                                              
\multicolumn {1}{c}{galaxy}  &  [MTT05]$^{a}$ & $\alpha_{2000}$$^{b}$ & $\delta_{2000}$$^{b}$ & $d$ [arcmin]$^{c}$ & $cz$ [km s$^{-1}$]  &  $i^{\prime}$$^{b}$  &  $r^{\prime}$-$i^{\prime}$$^{b}$  &  $M _{B}$  &  \multicolumn {1}{c}{type}   \\
\hline                                                                                                                                                                                                    
\endhead                                                                                                                                                                                                  
\hline                                                                                                                                                                                                    
\endfoot                                                                                                                                                                                                  
CGCG 20-39                   &      021       &      14 58 48.71      &      02 01 24.6       &       117.8        &     1800$\pm$30     &        13.25         &               0.39                &  -17.39   &              E               \\
NGC 5806                     &      037       &      15 00 00.39      &      01 53 28.7       &        98.7        &     1350$\pm$30     &        11.41         &               0.46                &  -19.03   &           SAB(s)b            \\
N5846\_55                    &      042       &      15 00 16.58      &      02 18 02.6       &       102.1        &     1830$\pm$30     &        14.19         &               0.36                &  -16.55   &              S0              \\
NGC 5811                     &      046       &      15 00 27.40      &      01 37 24.1       &        90.5        &     1529$\pm$15     &        13.83         &               0.31                &  -17.30   &            SB(s)m            \\
N5846\_53                    &      048       &      15 00 33.03      &      02 13 49.2       &        96.6        &     1260$\pm$30     &        16.03         &               0.27                &  -14.83   &             dE,N             \\
N5846\_47                    &      055       &      15 00 52.59      &      01 24 17.7       &        85.0        &     1890$\pm$60     &        15.35         &               0.35                &  -15.37   &            dE5,N             \\
N5846\_45                    &      058       &      15 00 59.36      &      01 52 36.2       &        84.1        &     2190$\pm$90     &        17.56         &               0.28                &  -13.32   &             dE,N             \\
N5846\_43                    &      059       &      15 00 59.36      &      01 38 57.1       &        82.5        &     2430$\pm$60     &        16.80         &               0.41                &  -13.84   &             dE1              \\
N5846\_49                    &      060       &      15 01 00.86      &      01 00 49.8       &        89.5        &     1740$\pm$30     &        17.48         &              -0.10                &  -14.15   &             dIrr             \\
N5846\_54                    &      061       &      15 01 03.11      &      00 42 27.4       &        97.7        &     1770$\pm$60     &        14.69         &               0.32                &  -16.20   &             S0/a             \\
N5846\_48                    &      063       &      15 01 06.96      &      02 05 25.2       &        85.7        &     1950$\pm$60     &        17.11         &               0.24                &  -13.74   &              dE              \\
NGC 5813                     &      064       &      15 01 11.23      &      01 42 07.1       &        79.7        &     1973$\pm$06     &        10.75         &               0.46                &  -19.63   &             E1-2             \\
N5846\_39                    &      068       &      15 01 15.33      &      01 29 53.5       &        78.8        &     2220$\pm$90     &        16.57         &               0.27                &  -14.62   &             dIrr             \\
N5846\_40                    &      069       &      15 01 15.89      &      01 46 24.5       &        79.0        &     1500$\pm$60     &        16.73         &               0.27                &  -14.24   &             dE5              \\
N5846\_37                    &      073       &      15 01 38.38      &      01 43 19.8       &        73.1        &     2280$\pm$60     &        16.47         &               0.40                &  -14.22   &            dE2,N             \\
N5846\_38                    &      075       &      15 01 38.61      &      01 52 12.6       &        74.4        &     2160$\pm$30     &        16.42         &               0.27                &  -14.62   &             dIrr             \\
UGC 9661                     &      083       &      15 02 03.50      &      01 50 28.6       &        67.9        &     1241$\pm$06     &        13.62         &               0.24                &  -17.55   &           SB(rs)d            \\
N5846\_29                    &      088       &      15 02 28.13      &      01 21 51.1       &        62.0        &     1470$\pm$90     &        17.11         &               0.25                &  -13.95   &              dE              \\
N5846\_30                    &      090       &      15 02 33.02      &      01 56 08.2       &        62.3        &     1650$\pm$60     &        17.07         &               0.24                &  -13.93   &              dE              \\
N5846\_31                    &      091       &      15 02 36.03      &      02 01 39.5       &        63.6        &     1980$\pm$60     &        16.87         &               0.33                &  -13.99   &             dE5              \\
N5846\_35                    &      113       &      15 03 44.28      &      02 33 08.1       &        70.2        &     1770$\pm$60     &        16.21         &               0.29                &  -14.65   &             dE1              \\
N5846\_26                    &      114       &      15 03 49.93      &      00 58 31.7       &        54.9        &     2010$\pm$90     &        15.88         &               0.07                &  -15.65   &             dIrr             \\
N5846\_25                    &      115       &      15 03 50.31      &      01 07 36.5       &        49.0        &     1590$\pm$30     &        14.85         &               0.39                &  -15.79   &             dE3              \\
N5846\_41/42$^{d}$           &       -        &      15 03 55.94      &      00 25 51.1       &        80.2        &    1600$\pm$300     &        15.80         &              -0.09                &  -16.23   &             dIrr             \\
NGC 5831                     &      122       &      15 04 06.99      &      01 13 11.7       &        42.4        &     1655$\pm$03     &        11.20         &               0.45                &  -19.26   &              E3              \\
N5846\_20                    &      124       &      15 04 08.45      &      01 31 28.0       &        35.5        &     1860$\pm$30     &        15.15         &               0.26                &  -15.99   &             dIrr             \\
N5846\_32                    &      125       &      15 04 13.07      &      02 32 34.6       &        65.7        &     1890$\pm$60     &        14.48         &               0.39                &  -16.22   &             dE2              \\
N5846\_23                    &      132       &      15 04 24.69      &      02 06 52.5       &        43.6        &     1770$\pm$30     &        16.05         &               0.32                &  -14.81   &            dE4               \\
N5846\_16                    &      142       &      15 04 42.90      &      01 17 27.2       &        32.6        &     1980$\pm$60     &        15.39         &               0.40                &  -15.31   &            dE1               \\
N5846\_18                    &      144       &      15 04 48.50      &      01 58 50.6       &        33.8        &     1950$\pm$30     &        17.09         &               0.19                &  -14.07   &             dIrr             \\
N5846\_14                    &      148       &      15 05 04.40      &      01 57 51.5       &        30.2        &     2368$\pm$05     &        16.54         &               0.14                &  -14.82   &             dIrr             \\
NGC 5838                     &      159       &      15 05 26.23      &      02 05 57.4       &        33.6        &     1358$\pm$09     &        10.67         &               0.49                &  -19.69   &             SA0              \\
NGC 5839                     &      160       &      15 05 27.48      &      01 38 05.3       &        15.5        &     1226$\pm$15     &        11.90         &               0.44                &  -18.57   &           SAB(rs)            \\
N5846\_06                    &      162       &      15 05 28.74      &      01 17 33.1       &        24.1        &     2308$\pm$05     &        17.65         &              -0.03                &  -14.04   &             dIrr             \\
N5846\_02                    &      165       &      15 05 31.83      &      01 35 15.4       &        14.4        &     0900$\pm$30     &        16.63         &               0.30                &  -14.23   &             dE2,N            \\
N5846\_05                    &      167       &      15 05 37.73      &      01 18 11.2       &        22.3        &     2040$\pm$60     &        16.02         &               0.34                &  -14.82   &            dE1               \\
N5846\_04                    &      177       &      15 05 50.57      &      01 54 29.7       &        20.6        &     1770$\pm$30     &        15.52         &               0.35                &  -15.19   &            dE1               \\
N5846\_09                    &      180       &      15 05 53.25      &      02 00 27.0       &        25.7        &     1290$\pm$30     &        17.33         &               0.12                &  -14.15   &             dIrr             \\
NGC 5845                     &      184       &      15 06 00.78      &      01 38 01.6       &        7.3         &     1451$\pm$09     &        11.78         &               0.47                &  -18.68   &              E               \\
N5846\_19                    &      187       &      15 06 03.32      &      02 11 05.5       &        35.4        &     1620$\pm$60     &        15.44         &               0.02                &  -16.00   &             dIrr             \\
N5846\_03                    &      191       &      15 06 06.73      &      01 19 20.8       &        17.9        &     2340$\pm$60     &        16.63         &               0.32                &  -14.18   &             dE4              \\
N5846\_12                    &      192       &      15 06 11.32      &      02 05 46.4       &        29.8        &     1799$\pm$05     &        16.01         &               0.05                &  -15.72   &             dIrr             \\
NGC 5846 A                   &      201       &      15 06 29.19      &      01 35 41.5       &        0.6         &     2200$\pm$15     &        13.56         &               0.41                &  -17.02   &            cE2-3             \\
NGC 5846                     &      202       &      15 06 29.28      &      01 36 20.2       &        0.0         &     1710$\pm$60     &        10.43         &               0.48                &  -19.96   &             E0-1             \\
N5846\_44                    &       -        &      15 06 34.25      &      00 12 55.7       &        83.4        &     2010$\pm$90     &        16.76         &               0.28                &  -14.26   &             dE,N             \\
N5846\_01                    &      205       &      15 06 34.27      &      01 33 31.6       &        03.1        &     1500$\pm$30     &        14.56         &               0.45                &  -15.90   &             dE1              \\
NGC 5848                     &      206       &      15 06 35.03      &      02 00 17.3       &        24.0        &     1260$\pm$30     &        12.96         &               0.40                &  -17.67   &              S0              \\
N5846\_50                    &       -        &      15 06 40.97      &      00 04 36.3       &        91.8        &     1710$\pm$30     &        16.79         &               0.28                &  -14.11   &             dE3              \\
N5846\_27                    &      212       &      15 06 42.03      &      00 38 03.7       &        58.4        &     2340$\pm$60     &        17.00         &               0.23                &  -14.26   &             dIrr             \\
NGC 5850                     &      233       &      15 07 07.68      &      01 32 39.2       &        10.3        &     2550$\pm$60     &        11.46         &               0.49                &  -18.87   &            SB(r)b            \\
N5846\_17                    &      241       &      15 07 34.18      &      01 07 13.1       &        33.3        &     1620$\pm$30     &        16.52         &               0.26                &  -14.54   &             dE4              \\
N5846\_13                    &      244       &      15 07 37.23      &      02 01 09.4       &        30.1        &     1890$\pm$60     &        14.61         &               0.38                &  -16.13   &            dE1,N             \\
NGC 5854                     &      246       &      15 07 47.69      &      02 34 07.0       &        61.0        &     1736$\pm$09     &        11.64         &               0.39                &  -19.09   &            SB(s)0            \\
N5846\_11                    &      247       &      15 07 47.82      &      01 17 31.4       &        27.2        &     2100$\pm$60     &        14.61         &               0.35                &  -16.22   &            dE2,N             \\
N5846\_22                    &      252       &      15 08 01.37      &      02 09 03.8       &        40.0        &     1080$\pm$30     &        16.57         &               0.35                &  -14.23   &             dE2              \\
N5846\_07                    &      256       &      15 08 05.61      &      01 39 05.7       &        24.2        &     1830$\pm$30     &        14.40         &               0.38                &  -16.16   &             S0/a             \\
N5846\_08                    &      260       &      15 08 09.25      &      01 36 29.7       &        25.0        &     2130$\pm$60     &        15.31         &               0.32                &  -15.53   &             dIrr             \\
N5846\_10                    &      261       &      15 08 12.38      &      01 29 58.8       &        26.5        &     1620$\pm$30     &        16.69         &               0.36                &  -14.10   &            dE1,N             \\
N5846\_15                    &      266       &      15 08 22.69      &      01 47 54.9       &        30.6        &     1680$\pm$60     &        16.72         &               0.27                &  -14.20   &            dE4,N             \\
N5846\_21                    &      276       &      15 08 47.18      &      01 53 59.8       &        38.7        &     2010$\pm$60     &        15.37         &               0.32                &  -15.50   &             dE6              \\
N5846\_28                    &      283       &      15 09 04.29      &      00 49 19.1       &        60.9        &     1649$\pm$05     &        15.60         &               0.30                &  -15.34   &             dIrr             \\
N5846\_33                    &      287       &      15 09 07.86      &      00 43 29.5       &        66.1        &     1680$\pm$30     &        16.64         &               0.24                &  -14.43   &             dIrr             \\
N5846\_24                    &      290       &      15 09 14.97      &      01 55 17.1       &        45.5        &     1740$\pm$30     &        15.40         &               0.31                &  -15.44   &             dE4              \\
NGC 5864                     &      299       &      15 09 33.56      &      03 03 09.9       &        98.3        &     1800$\pm$30     &        11.42         &               0.40                &  -19.21   &           SB(s)sp            \\
NGC 5869                     &       -        &      15 09 49.41      &      00 28 12.2       &        84.5        &     1934$\pm$15     &        11.48         &               0.39                &  -19.01   &              S0              \\
UGC 9746                     &      305       &      15 10 16.54      &      01 56 03.3       &        60.1        &     1830$\pm$60     &        13.64         &               0.33                &  -17.24   &             Sbc              \\
UGC 9751                     &      311       &      15 10 58.44      &      01 26 15.9       &        68.0        &     1574$\pm$15     &        15.31         &               0.26                &  -15.80   &             Scd              \\
N5846\_34                    &      313       &      15 11 01.33      &      01 40 50.1       &        68.2        &     1710$\pm$30     &        16.03         &               0.32                &  -14.81   &            dE2               \\
N5846\_36                    &      317       &      15 11 21.63      &      01 36 37.6       &        73.1        &     1950$\pm$90     &        16.27         &               0.21                &  -14.70   &              dE              \\
UGC 9760                     &      321       &      15 12 02.18      &      01 41 51.3       &        83.4        &     2024$\pm$03     &        14.85         &               0.38                &  -16.15   &              Sd              \\
N5846\_46                    &      323       &      15 12 08.15      &      01 35 08.6       &        84.7        &     2010$\pm$60     &        15.85         &               0.28                &  -15.25   &             dE1              \\
N5846\_52                    &       -        &      15 12 24.05      &      02 04 48.2       &        93.1        &     1740$\pm$30     &        15.69         &               0.32                &  -15.17   &            dE3,N             \\
N5846\_56                    &       -        &      15 12 31.74      &      00 48 45.3       &       102.3        &     1830$\pm$60     &        15.14         &               0.32                &  -15.82   &             dIrr             \\
N5846\_51                    &       -        &      15 12 41.39      &      01 37 23.7       &        93.0        &     1920$\pm$60     &        16.16         &               0.28                &  -14.80   &             dE4              \\
\end{longtable}
\begin {flushleft}
$^{a}$ galaxy identification from \citet{mahdavi}           \\
$^{b}$ coordinates and magnitudes taken from SDSS DR4       \\
$^{c}$ projected radial distance to NGC 5846                \\
$^{d}$ N5846\_41/42 are two HII regions classified as individual galaxies in SDSS (see section \ref{individual}         \\
\end{flushleft}
}

\end{document}